# The intermittent excitation of geodesic acoustic mode by nonlinear Instanton of electron drift wave envelope in L-mode discharge near tokamak edge

Z. Y. Liu [1*], Y. Z. Zhang [2], S. M. Mahajan [3], A. D. Liu [1], C. Zhou [1], T. Xie [4]

[1] School of Nuclear Science and Technology, University of Science and Technology of China, Hefei, Anhui 230026, People's Republic of China

[2] Center for Magnetic Fusion Theory, Chinese Academy of Sciences, Hefei, Anhui 230026, People's Republic of China

[3] Institute for Fusion Studies, University of Texas at Austin, Austin, Texas 78712, USA

[4] School of Science, Sichuan University of Science and Engineering, Zigong, Sichuan 643000, People's Republic of China

[*]E-mail of corresponding author: lzy0928@mail.ustc.edu.cn

## Abstract

There are two distinct phases in the evolution of drift wave envelope in the presence of zonal flow. A long-lived standing wave phase, which we call the Caviton, and a short-lived traveling wave phase (in radial direction) we call the Instanton. For drift wave turbulence driven by ion temperature gradient mode (ITG), these two stages of dynamics were displayed in [Zhang Y Z, Liu Z Y, Xie T, Mahajan S M and Liu J 2017 Physics of Plasmas **24** 122304]. In this paper we show that the dynamical attributes of ITG turbulence are readily replicated when the turbulence rotates in the electron direction; our model calculation deals specifically with the toroidal electron drift waves (EDW) in the well-known $\delta_e$ model. While the basic calculations are presented in parallel to the ITG counterpart, more emphasis is laid here on the motion of Instanton; several abrupt phenomena observed in tokamaks, such as intermittent excitation of geodesic acoustic mode (GAM) shown in this paper, could be attributed to the sudden and fast radial motion of Instanton. The composite drift wave – zonal flow system evolves at the two well - separate scales: the micro and the meso-scale. The eigenmode equation of the model defines the zero order (micro-scale) variation; it is solved by making use of the two dimensional (2D) weakly asymmetric ballooning theory (WABT), a theory suitable for modes localized to rational surface like drift waves, and then refined by shifted inverse power method, an iterative finite difference method. The next order is the equation of EDW envelope (containing group velocity of EDW) which is modulated by the zonal flow generated by Reynolds stress of EDW. This equation is coupled to the zonal flow equation, and numerically solved in spatiotemporal representation; the results are displayed in self-explanatory graphs. One observes a strong correlation between the Caviton-Instanton transition and the zero-crossing of radial group velocity of EDW. The calculation brings out the defining characteristics of the Instanton: it begins as a linear traveling wave right after the transition. Then, it evolves to a nonlinear stage with increasing frequency all the way to 20 kHz. The modulation to Reynolds stress in zonal flow equation will cause resonant excitation to GAM. The intermittency is shown due to the random phase mixing between multiple central rational surfaces in the reaction region.

**Keywords**: zonal flow, drift wave, Caviton, Instanton, L-H mode

## 1. Introduction

Zonal flows are ubiquitous in nature and in the laboratory [1,2]. In a plasma torus like tokamak they appear in two distinct manifestations: the toroidal low frequency zonal flow (TLFZF) and GAM. The latter, usually, are observed as 'intermittent excitation' [2], that does not persist. Though the existence of intermittent excitations has been reported in many simulations [3-7], no physical mechanism has been clearly identified. There are many theories for GAM onset, but none of them would lead to an intermittent excitation. In this paper we propose and hopefully show that the Caviton to Instanton transition may trigger the 'intermittent excitation'; these two already defined objects [8] constitute two distinct phases in the evolution of drift wave envelope in the presence of zonal flow. In the Caviton phase, the drift wave envelope is slowly growing and deforming; there is no radial propagation. However when the radial group velocity (the velocity of EDW energy flow) passes through zero (zero-crossing), it ushers in the transition from the Caviton into the Instanton; the latter moves rapidly beyond the radial boundary. Downward (upward) zero-crossing of radial group velocity induces inward (outward) motion of the Instanton (see Figure 7 of [8] and Figure 10 of this paper). While the two distinct patterns of drift wave envelope have been introduced and explained in [8], we intend to begin this paper by further dwelling on, and elaborating the Caviton-Instanton dynamics for completeness and clarity. Several essential details can be seen in the last paragraph of this section.

One must confront at the outset that the best-known drift waves in the electron direction – namely the collisionless trapped electron mode (CTEM or simply TEM) and dissipative trapped electron mode (DTEM) [9,10] – are not likely to be major constituents of the L-mode turbulence because of high collisionality in the edge region where GAM activity is observed. In this paper, we will resort to the instability associated with the relatively simple but well-known $\delta_e$-model [11-13]; it will be examined as a candidate for edge turbulence. We choose this model for its relative simplicity even though more realistic models, such as impurity driven drift waves, do exist in literature [14]. Notice that the description of the latter requires the simultaneous treatment of several fields: the perturbed density, temperature and parallel velocities for electrons, main ions and impurities. A two-dimensional (2D) theory (for example, in the 2D ballooning representation) for such a system is quite complicated. No doubt that these complexities could be significant, but a first attempt with a simpler model is quite appropriate. We expect that the fuller models may be essential to predict linear growth rates and marginal stability criteria, but quantities like Reynolds stress and group velocity (determined by the structure of the mode potential) may be well approximated in the simpler model. The last but not the least reason for the simpler approach is that a lack of experimental knowledge on impurities (at the L-H transition threshold) would be a serious impediment to the analysis of the impurity driven mode. For brevity, the toroidal electron drift wave in $\delta_e$-model will be called $\delta_e$-mode.

In this paper, the $\delta_e$-mode will be treated as a 2D system. In order to obtain the 2D structure of the mode it is important to keep all the translational symmetry breaking (TSB) terms [13,15]. When TSB terms are neglected, the resulting 1D equation yields two branches: the slab-like Pearlstein-Berk branch [16], stabilized by the magnetic shear [17,18], and the Chen-Cheng branch [12,19], an intrinsic toroidal branch that disappears in the slab limit. For the latter branch the shear stabilization is greatly reduced owing to 'tunneling' [12]; consequently the finite non-adiabatic part of electron density can render the mode unstable. The growth rate of the latter branch is, likely, greater than the former one for same $\delta_e$. Therefore, in this paper $\delta_e$-mode solely refers to the intrinsic toroidal branch.

The remaining part of the paper is organized as follows. In section 2 the multiple scale derivative expansion method in spatiotemporal configuration is applied to $\delta_e$-mode; the zero order (micro-scale) equation defines the relevant drift wave. The first order (meso-scale) yields an equation for drift wave envelope modulated by the zonal flow. In section 3 the micro-scale equation is solved by making use of the 2D WABT. The solution is then chosen to be the initial guess of an iterative finite difference method to acquire more accurate results in section 4. Use is made of the two types of solutions to obtain the spatiotemporal structure of Reynolds stress and group velocity in section 5. This sets the stage for substituting the principal ingredients into the zonal flow - drift wave envelope set equations (37-38), which does not contain the high frequency branch – GAM in zonal flow equation. The information regarding Instanton is given in section 6. In sub-section 6.1 the correlation between occurrence of spikes and zero-crossing of radial group velocity is highlighted. The Caviton to Instanton transition is displayed in sub-section 6.2 by a set of snapshot in the movie. In sub-section 6.3 a few interesting properties are presented: (a) Instanton is the linear-to-nonlinear traveling wave shown by analytic solution near zero-crossing; (b) The nonlinear Instanton keeps growing then falls off and disappears; (c) the Instanton frequency grows with its amplitude, may reach the frequency range of GAM. Section 7 begins with introducing the zonal flow equation containing the high frequency branch – GAM for EDW in existing literature. It is followed by two sub-sections. The causal relationship of GAM excitation to Instanton frequency is depicted in sub-section 7.1 for GAM pertaining to single central rational surface. In sub-section 7.2, the intermittency is explained by nonlinear coupling between multiple central rational surfaces. We will show that this coupling introduces the random phase mixing between multiple central rational surfaces in the reaction region, which leads to the intermittency of GAM excitation. Discussions and conclusions are given in section 8. In Appendix

A, we derive the 'extended' $\delta_e$-model by including TSB terms in $(x,l)$ representation. The conditions for a valid WABT are discussed in Appendix B. In spatiotemporal representation, the zonal flow equation containing both high-low frequency components in the EDW system is given in Appendix C; equivalent to the same equations in Fourier representation in existing literature. The normalization and numerical methods for the zonal flow equation set are depicted in Appendix D.

## 2. Multiple scale derivative expansion method for $\delta_e$-mode

The $\delta_e$-mode of this paper is built on a large-aspect-ratio, up-down symmetric tokamak equilibrium with concentric circular magnetic surfaces. We start from the first two linear moment equations – the continuity equation of warm ion fluid, and ion parallel momentum equation under modulation of zonal flow

$$\left(\frac{\partial}{\partial t}+\overline{\boldsymbol{v}}\cdot\nabla\right)\left(1-i\delta_e-\rho_s^2\nabla_\perp^2\right)\varphi+i\widehat{\omega}_{*i}\rho_s^2\nabla_\perp^2\varphi+i\widehat{\omega}_{*e}\varphi-2i\left(1+\delta_\tau\right)\widehat{\omega}_{de}\varphi+\nabla_\parallel u_\parallel=0, \quad (1)$$

$$\left(\frac{\partial}{\partial t}+\overline{\boldsymbol{v}}\cdot\nabla\right)u_\parallel+\left(1+\delta_\tau\right)c_s^2\nabla_\parallel\varphi=0, \quad (2)$$

where the perturbed density is defined by $(1-i\delta_e)\varphi$, $\delta_e$ represents the non-adiabatic response of electron density, a small number in this paper, $\overline{\boldsymbol{v}}\equiv\rho_s c_s \boldsymbol{b}\times\nabla\overline{\varphi}$ is the zonal flow, $\varphi$ ($\overline{\varphi}$) is the dimensionless perturbed (zonal) electrostatic potential normalized to electron thermal energy, $u_\parallel$ is the perturbed ion parallel velocity, $\rho_s\equiv\sqrt{m_i T_e^{(0)}}/eB$ is the ion Larmor radius at electron temperature, $c_s\equiv\sqrt{T_e^{(0)}/m_i}$ is the ion sound speed, $\widehat{\omega}_{*e}\equiv i\rho_s c_s \boldsymbol{b}\times\nabla\ln n^{(0)}\cdot\nabla$ ($\widehat{\omega}_{*i}=\overline{\eta}_i\widehat{\omega}_{*e}$) is the electron (ion) diamagnetic frequency operator, $\widehat{\omega}_{de}\equiv -i\rho_s c_s \boldsymbol{\kappa}\times\boldsymbol{b}\cdot\nabla$ is the curvature frequency operator, $\boldsymbol{b}$ is the unit vector along the equilibrium magnetic field, $\boldsymbol{\kappa}$ is the magnetic curvature. Warm ion effect is considered here, $\bar{\eta}_i \equiv \tau_i(1+\eta_i)$, $\eta_i \equiv d\ln T_i^{(0)}/d\ln n^{(0)}$, $\tau_i \equiv T_i^{(0)}/T_e^{(0)}$, where $n^{(0)}$ is the equilibrium density, $T_e^{(0)}(T_i^{(0)})$ is the equilibrium electron (ion) temperature. $\delta_\tau \equiv \tau_i(1-i\delta_e)$.

The methodology of [8] begins with the multiple scale expansion method in which the derivatives take the form [20]

$$\frac{\partial}{\partial t} \to \frac{\partial}{\partial \tilde{t}} + \varepsilon_E \frac{\partial}{\partial t} \to -i\omega + \varepsilon_E \frac{\partial}{\partial t}, \quad \frac{\partial}{\partial \boldsymbol{r}} \to \frac{\partial}{\partial \tilde{\boldsymbol{r}}} + \varepsilon_E \frac{\partial}{\partial \boldsymbol{r}}, \quad \varphi(t,\boldsymbol{r}) \to \bar{\phi}(t,\boldsymbol{r})\varphi(\tilde{t},\tilde{\boldsymbol{r}}), \quad (3)$$

where $\tilde{t}, \tilde{\boldsymbol{r}}$ ($t,\boldsymbol{r}$) denotes fast (slow) scale variables (in toroidal coordinate $\boldsymbol{r} \equiv (r,\vartheta,\zeta)$, corresponding to the radial, poloidal, and toroidal directions respectively), $\varepsilon_E \ll 1$ is introduced for bookkeeping. The $\delta_e$-mode potential $\varphi(\tilde{t},\tilde{\boldsymbol{r}})$ varies on fast scale (usually with high frequency $\omega$: $\varphi(\tilde{t},\tilde{\boldsymbol{r}}) = \varphi(\tilde{\boldsymbol{r}})\exp(-i\omega\tilde{t})$), while the drift wave envelope $\bar{\phi}(t,\boldsymbol{r})$ which is modulated by zonal flow $\bar{\upsilon}$ has slow variation. The zonal flow in slow scale is treated to be small quantity, $\bar{\boldsymbol{\upsilon}} \cdot \nabla \to \varepsilon_E \bar{\upsilon}(\partial/\partial \tilde{y})$. Here $y \equiv r_j \vartheta$ and $\partial/\partial \tilde{y} \to -i(m+l)/r_j \approx -ik_\vartheta$, $k_\vartheta \equiv m/r_j$ (negative in this paper), $m$ ($l$) is the central (sideband) poloidal number and $r_j$ denotes the position of rational surface. The differential operators of interest, calculated to order $\varepsilon_E$, are

$$\nabla_\perp^2 = \frac{\partial^2}{\partial \tilde{r}^2} - k_\vartheta^2 + 2\varepsilon_E\left(\frac{\partial}{\partial r}\frac{\partial}{\partial \tilde{r}} - ik_\vartheta \frac{\partial}{\partial y}\right) \equiv \nabla_{\perp,0}^2 + 2\varepsilon_E\left(\frac{\partial}{\partial r}\frac{\partial}{\partial \tilde{r}} - ik_\vartheta \frac{\partial}{\partial y}\right), \quad (4)$$

$$\hat{\omega}_{*e} = -i\frac{\rho_s c_s}{L_n}\left(-ik_\vartheta + \varepsilon_E \frac{\partial}{\partial y}\right) \equiv \hat{\omega}_{*e,0} + \varepsilon_E \hat{\omega}_{*e,1}, \quad \hat{\omega}_{*i} = \hat{\omega}_{*i,0} + \varepsilon_E \hat{\omega}_{*i,1}, \quad (5)$$

$$\hat{\omega}_{de} = -i\frac{\rho_s c_s}{R}\left[\sin\vartheta\left(\frac{\partial}{\partial \tilde{r}} + \varepsilon_E \frac{\partial}{\partial r}\right) + \cos\vartheta\left(-ik_\vartheta + \varepsilon_E \frac{\partial}{\partial y}\right)\right] \equiv \hat{\omega}_{de,0} + \varepsilon_E \hat{\omega}_{de,1}, \quad (6)$$

and

$$\nabla_\parallel u = \left[1 + \varepsilon_E \frac{1}{i\omega}\left(\frac{\partial}{\partial t} - ik_\vartheta \bar{\upsilon}\right)\right](1+\delta_\tau)\frac{c_s^2}{i\omega}\nabla_\parallel^2 \varphi. \quad (7)$$

Substituting equations (4-7) into equation (1), yields

$$\left[-i\omega + \varepsilon_E\left(\frac{\partial}{\partial t} - ik_g\bar{\upsilon}\right)\right]\left\{1 - i\delta_e - \rho_s^2\left[\nabla_{\perp,0}^2 + 2\varepsilon_E\left(\frac{\partial}{\partial r}\frac{\partial}{\partial \tilde{r}} - ik_g\frac{\partial}{\partial y}\right)\right]\right\}\varphi$$

$$+i\rho_s^2\left(\hat{\omega}_{*i,0} + \varepsilon_E\hat{\omega}_{*i,1}\right)\left[\nabla_{\perp,0}^2 + 2\varepsilon_E\left(\frac{\partial}{\partial r}\frac{\partial}{\partial \tilde{r}} - ik_g\frac{\partial}{\partial y}\right)\right]\varphi + i\left(\hat{\omega}_{*e,0} + \varepsilon_E\hat{\omega}_{*e,1}\right)\varphi \quad . \quad (8)$$

$$-2i(1+\delta_\tau)\left(\hat{\omega}_{de,0} + \varepsilon_E\hat{\omega}_{de,1}\right)\varphi + \left[-i\omega + \varepsilon_E\left(\frac{\partial}{\partial t} - ik_g\bar{\upsilon}\right)\right](1+\delta_\tau)\frac{c_s^2}{i\omega}\nabla_\parallel^2\varphi = 0$$

For the multiple scale wave function $\varphi(t,\mathbf{r}) \to \bar{\phi}(t,\mathbf{r})\varphi(\tilde{\mathbf{r}})$ (assuming stationary background micro-turbulence), the zeroth order equation

$$\left[\left(1 + \frac{\hat{\omega}_{*i,0}}{\omega}\right)\rho_s^2\nabla_{\perp,0}^2 - (1-i\delta_e) + \frac{\hat{\omega}_{*e,0}}{\omega} - 2(1+\delta_\tau)\frac{\hat{\omega}_{de,0}}{\omega} - (1+\delta_\tau)\frac{c_s^2}{\omega^2}\nabla_\parallel^2\right]\varphi(\tilde{\mathbf{r}}) = 0 \quad (9)$$

corresponds to equation (A1) in Appendix A. From the first order equation

$$\frac{\partial\bar{\phi}(t,r,\vartheta)}{\partial t} + \upsilon_{gr}(\vartheta)\frac{\partial\bar{\phi}(t,r,\vartheta)}{\partial r} + \upsilon_{gy}(\vartheta)\frac{\partial\bar{\phi}(t,r,\vartheta)}{r_j\partial\vartheta} = ik_g\bar{\upsilon}(t,r)\bar{\phi}(t,r,\vartheta), \quad (10)$$

we calculate the radial group velocity

$$\upsilon_{gr}(\vartheta) \equiv -\frac{2(\omega - \upsilon_*\bar{\eta}_i k_g)\rho_s^2\langle K_r\rangle + 2(1+\delta_\tau)\frac{\rho_s c_s}{R}\sin\vartheta}{1 - i\delta_e + \rho_s^2 k_g^2 + \rho_s^2\langle K_r^2\rangle - (1+\delta_\tau)\frac{c_s^2}{\omega^2}\langle\nabla_\parallel^2\rangle}, \quad (11)$$

and the poloidal group velocity

$$\upsilon_{gy}(\vartheta) \equiv \frac{\upsilon_*\left[1 - \bar{\eta}_i\rho_s^2\left(\langle K_r^2\rangle + 3k_g^2\right)\right] + 2\omega\rho_s^2 k_g - 2(1+\delta_\tau)\frac{\rho_s c_s}{R}\cos\vartheta}{1 - i\delta_e + \rho_s^2 k_g^2 + \rho_s^2\langle K_r^2\rangle - (1+\delta_\tau)\frac{c_s^2}{\omega^2}\langle\nabla_\parallel^2\rangle}, \quad (12)$$

where $K_r \equiv -i(\partial/\partial\tilde{r})$, $\langle...\rangle \equiv \int d\tilde{\mathbf{r}}\varphi^*(\tilde{\mathbf{r}})...\varphi(\tilde{\mathbf{r}})/\int d\tilde{\mathbf{r}}\varphi^*(\tilde{\mathbf{r}})\varphi(\tilde{\mathbf{r}})$ denotes average over the radial fast scale $\tilde{r}$.

Before ending this section it is appropriate to point out that the explicitly different notation for two different scales makes sense only for derivative expansion in a singular perturbation theory. Such distinctions ($\tilde{r}$, $\tilde{t}$) will not appear in the rest of paper.

## 3. WABT for solving $\delta_e$-mode

In the toroidal coordinates $(r,\vartheta,\zeta)$ the 2D mode can be expressed in the $(x,l)$ representation near the rational surface $r_j$

$$\varphi(r) \equiv \varphi(r,\vartheta,\zeta) = \exp[i(n\zeta - m\vartheta)]\sum_l \varphi_l(x)\exp(-il\vartheta), \qquad (13)$$

where $n$ is the toroidal mode number (a good quantum number in this paper), $m = nq(r_j)$ is the poloidal mode number, $q$ is the safety factor of tokamak. The rational surface $r_j$ is called central rational surface, defined by the pair of integer $(n,m)$ and monotonic $q(r)$. The extended $\delta_e$-model [11-13], including the translational symmetric breaking (TSB) terms, comes from the zeroth order equation (9). In $(x,l)$ representation, where $x \equiv k_\vartheta \hat{s}(r-r_j)$ and $\hat{s} \equiv d\ln q(r_j)/d\ln r$ is the magnetic shear, it reads

$$\hat{k}_\vartheta^2 \hat{s}^2 \frac{\hat{\omega}}{\hat{\omega}_s}\left(1+\frac{\bar{\eta}_i}{\hat{\omega}}\right)\frac{\partial^2 \varphi_l}{\partial x^2} - \frac{\hat{\omega}}{\hat{\omega}_s}\left(1-i\delta_e + \hat{k}_\vartheta^2 - \frac{1}{\hat{\omega}} + \frac{\bar{\eta}_i}{\hat{\omega}}\hat{k}_\vartheta^2\right)\varphi_l + (1+\delta_\tau)\frac{\hat{\omega}_s}{\hat{\omega}}(x-l)^2 \varphi_l$$

$$-(1+\delta_\tau)\frac{\hat{\omega}_{de}}{\hat{\omega}_s}\left[\hat{s}\frac{\partial}{\partial x}(\varphi_{l+1}-\varphi_{l-1}) + (\varphi_{l+1}+\varphi_{l-1})\right]$$

$$+\left\{-\frac{\hat{\omega}}{\hat{\omega}_s}\left[2\hat{k}_\vartheta^2 + (1-i\delta_e)\frac{1}{\hat{s}}\frac{r_j}{L_{T_e}}\right]\varphi_l + \hat{k}_\vartheta^2 \frac{\bar{\eta}_i}{\hat{\omega}_s}\left[-3-\frac{1}{\hat{s}}\frac{r_j}{L_n}+\left(1+\frac{1}{\hat{s}}\frac{r_j}{L_n}\right)\hat{s}^2\frac{\partial^2}{\partial x^2}\right]\varphi_l\right.$$

$$\left.+\frac{1}{\hat{\omega}_s}\left(1+\frac{1}{\hat{s}}\frac{r_j}{L_n}\right)\varphi_l - (1+\delta_\tau)\frac{\hat{\omega}_{de}}{\hat{\omega}_s}(\varphi_{l+1}+\varphi_{l-1})\right\}\left(\frac{l}{m}\right)$$

$$+\left\{\frac{1}{\hat{\omega}_s}\frac{1}{\hat{s}}\frac{r_j}{L_n}\left(1+\frac{1}{\hat{s}}\frac{r_j}{L_n}\right)\varphi_l - \frac{\hat{\omega}}{\hat{\omega}_s}\left[\hat{k}_\vartheta^2 + (1-i\delta_e)\frac{1}{\hat{s}^2}\frac{r_j^2}{L_{T_e}^2}\right]\varphi_l\right.$$

$$\left.+\hat{k}_\vartheta^2 \frac{\bar{\eta}_i}{\hat{\omega}_s}\left[-3-\frac{1}{\hat{s}}\frac{r_j}{L_n}\left(3+\frac{1}{\hat{s}}\frac{r_j}{L_{P_i}}\right)+\frac{1}{\hat{s}}\frac{r_j}{L_n}\left(1+\frac{1}{\hat{s}}\frac{r_j}{L_{P_i}}\right)\hat{s}^2\frac{\partial^2}{\partial x^2}\right]\varphi_l\right\}\left(\frac{l}{m}\right)^2 = 0 \qquad (14)$$

Equation (14) is derived in Appendix A, where $\hat{k}_\vartheta \equiv \rho_{s,j} k_\vartheta$, $\hat{\omega} \equiv \omega/\omega_{*e,j}$, $\hat{\omega}_s \equiv c_{s,j}/qR\omega_{*e,j}$, $\hat{\omega}_{de} \equiv \omega_{de,j}/\omega_{*e,j}$, subscript $j$ denotes equilibrium quantities on rational surface $r_j$, the electron (ion) diamagnetic frequency $\omega_{*e} \equiv -k_\vartheta T_e^{(0)}/eBL_n$ ($\omega_{*i} = \bar{\eta}_i \omega_{*e}$) and the curvature frequency $\omega_{de} \equiv -k_\vartheta T_e^{(0)}/eBR$, $R$ is the major radius, $B$ is the equilibrium magnetic field, $L_n$, $L_{T_e}$ and $L_{P_i}$ are the density, electron temperature and ion pressure gradient length respectively, and $e$ is the unit electric charge. The remaining part of this section will be written in such a manner that is more accessible to readers who are not familiar with WABT and/or the 2D ballooning theory; some known facts may need to be repeated for better readability.

For a high $n$ local mode pertaining to a given rational surface $r_j$, the monotonic safety factor $q(r)$ can be expanded up to the first order: $q(r) \approx q(r_j) + (dq/dr)(r-r_j) \equiv q(r_j) + x/n$. Then, one may develop the 2D theory by invoking the 2D Fourier-ballooning transform [13,15,21-23]

$$\varphi_l(x) = \frac{1}{2\pi}\int_{-\pi}^{\pi} d\lambda \int_{-\infty}^{+\infty} dk\, e^{ik(x-l)-i\lambda l}\varphi(k,\lambda). \tag{15}$$

For WABT, we may assume $\varphi(k,\lambda) := \psi(\lambda)\chi(k,\lambda)$, where $\psi(\lambda)$ is a fast varying function in $\lambda$, known as Floquet phase distribution (FPD), $\chi(k,\lambda)$ is the solution of ballooning equation (parameterized by $\lambda$ through $\sin\lambda$ and $\cos\lambda$). $\psi(\lambda)$ is localized around some $\lambda_*$. In the limit $\psi(\lambda) \to \delta(\lambda - \lambda_*)$ ($\delta$ denotes Dirac delta function), equation (15) reduces to the Lee-Van Dam representation [24]. Notice that equation (15) is the mathematical transform (the inverse transform does exist and unique). The 2D ballooning equation for the $\delta_e$-model in WABT is expanded to the second order of $\varepsilon_B \equiv (1/n)(d/d\lambda)$

$$\left[L_0(k,\lambda;\hat{\omega}) + \frac{iL_1(k,\lambda;\hat{\omega})}{n}\frac{\partial}{\partial\lambda} + \frac{L_2(k,\lambda;\hat{\omega})}{n^2}\frac{\partial^2}{\partial\lambda^2} + \cdots - \Omega(\hat{\omega})\right]\varphi(k,\lambda) = 0, \tag{16}$$

with

$$\Omega(\hat{\omega}) \equiv -\frac{\hat{\omega}}{\hat{\omega}_s}\left(1 - i\delta_e + \hat{k}_\vartheta^2 - \frac{1}{\hat{\omega}} + \frac{\bar{\eta}_i}{\hat{\omega}}\hat{k}_\vartheta^2\right), \tag{17}$$

$$L_0(k,\lambda;\hat{\omega}) = (1+\delta_\tau)\frac{\hat{\omega}_s}{\hat{\omega}}\frac{\partial^2}{\partial k^2} + \frac{\hat{\omega}}{\hat{\omega}_s}\left(1+\frac{\bar{\eta}_i}{\hat{\omega}}\right)\hat{k}_g^2\hat{s}^2 k^2$$
$$+2(1+\delta_\tau)\frac{\hat{\omega}_{de}}{\hat{\omega}_s}\left[\hat{s}k\sin(k+\lambda)+\cos(k+\lambda)\right] \quad (18)$$

$$L_1(k,\lambda;\hat{\omega}) = \frac{1}{q}\left\{\hat{k}_g^2\frac{\bar{\eta}_i}{\hat{\omega}_s}\left[-3-\frac{1}{\hat{s}}\frac{r_j}{L_n}-\left(1+\frac{1}{\hat{s}}\frac{r_j}{L_n}\right)\hat{s}^2 k^2\right]+\frac{1}{\hat{\omega}_s}\left(1+\frac{1}{\hat{s}}\frac{r_j}{L_n}\right)\right.$$
$$\left.-\frac{\hat{\omega}}{\hat{\omega}_s}\left[2\hat{k}_g^2+(1-i\delta_e)\frac{1}{\hat{s}}\frac{r_j}{L_{T_e}}\right]-2(1+\delta_\tau)\frac{\hat{\omega}_{de}}{\hat{\omega}_s}\cos(k+\lambda)\right\} \quad (19)$$

$$L_2(k,\lambda;\hat{\omega}) = \frac{1}{q^2}\left\{\hat{k}_g^2\frac{\bar{\eta}_i}{\hat{\omega}_s}\left[-3-\frac{1}{\hat{s}}\frac{r_j}{L_n}\left(3+\frac{1}{\hat{s}}\frac{r_j}{L_{P_i}}\right)-\frac{1}{\hat{s}}\frac{r_j}{L_n}\left(1+\frac{1}{\hat{s}}\frac{r_j}{L_{P_i}}\right)\hat{s}^2 k^2\right]\right.$$
$$\left.+\frac{1}{\hat{\omega}_s}\frac{1}{\hat{s}}\frac{r_j}{L_n}\left(1+\frac{1}{\hat{s}}\frac{r_j}{L_n}\right)-\frac{\hat{\omega}}{\hat{\omega}_s}\left[\hat{k}_g^2+(1-i\delta_e)\frac{1}{\hat{s}^2}\frac{r_j^2}{L_{T_e}^2}\right]\right\} \quad (20)$$

In addition to small $\varepsilon_B$, WABT requests another small parameter $\Xi \equiv \bar{L}_1/2\bar{L}_2$. This issue will be discussed in Appendix B.

The lowest order of equation (16)

$$\left[L_0(k,\lambda;\hat{\omega})-\Omega(\lambda)\right]\chi(k,\lambda)=0 \quad (21)$$

is, traditionally, called the ballooning equation with $\Omega(\lambda)$ representing the ($\lambda$-parameterized) effective local eigenvalue. The ballooning equation has two salient features: (a) its counter-part in the $(x,l)$ representation (calculated after removing all TSB terms from equation (A4) such as $l/m \to 0$ and $f(x) \to f(0)$) is translational invariant under the transform $(x,l) \to (x+1,l+1)$, and (b) the combined parity (CP) conservation, exhibited in $L_0(k,\lambda;\hat{\omega}) = L_0(-k,-\lambda;\hat{\omega})$ forces the "eigenvalue" $\Omega(\lambda)$ to be an even function of $\lambda$.

Substituting the solution of equation (21) into equation (16), and taking average over the first ballooning variable $k$, yields the differential equation governing $\psi(\lambda)$ (the 2D system is being solved in a series of two 1D equations)

$$\left[\frac{i\bar{L}_1(\lambda;\hat{\omega})}{n}\frac{d}{d\lambda}+\frac{\bar{L}_2(\lambda;\hat{\omega})}{n^2}\frac{d^2}{d\lambda^2}+\cdots\Omega(\lambda)-\Omega(\hat{\omega})\right]\psi(\lambda)=0, \tag{22}$$

where $\bar{L}_s$ ($s=1,2,...$) are $k$-averages of $L_{1,2}(k,\lambda;\hat{\omega})$.

For the WABT approximation, the Floquet phase function obeys

$$\frac{d^2\psi(\lambda)}{d\lambda^2}+P(\lambda)\frac{d\psi(\lambda)}{d\lambda}+Q(\lambda)\psi(\lambda)=0, \tag{23}$$

with

$$P(\lambda)\equiv\frac{n}{\bar{L}_2^{(0)}(\lambda;\hat{\omega})}\left[i\bar{L}_1^{(0)}(\lambda;\hat{\omega})+\frac{2\bar{L}_2^{(1)}(\lambda;\hat{\omega})}{n}\right], \tag{24}$$

$$Q(\lambda)\equiv\frac{n^2}{\bar{L}_2^{(0)}(\lambda;\hat{\omega})}\left[\Omega(\lambda)-\Omega(\hat{\omega})+\frac{i\bar{L}_1^{(1)}(\lambda;\hat{\omega})}{n}+\frac{\bar{L}_2^{(2)}(\lambda;\hat{\omega})}{n^2}\right], \tag{25}$$

$$\bar{L}_s^{(j)}(\lambda;\hat{\omega})\equiv\frac{\int_{-\infty}^{\infty}dk\,\chi^*(k,\lambda)L_s(k,\lambda;\hat{\omega})\frac{\partial^j\chi(k,\lambda)}{\partial\lambda^j}}{\int_{-\infty}^{\infty}dk\,\chi^*(k,\lambda)\chi(k,\lambda)}\ (s=1,2,\ j=1,2). \tag{26}$$

For the numerical calculation, we will use the edge parameters corresponding to an L-mode discharge in the DIII-D tokamak [25], the basic equilibrium parameters are listed in Table 1 (where $R$ ($a$) is the major (minor) radius of tokamak, $L_n$ ($L_{T_e}$) is the density (electron temperature) gradient length, $B$, $q$, $\hat{s}$, $n_e$ and $T_e$ are the equilibrium magnetic field, safety factor, magnetic shear, electron density and temperature at the position of rational surface $r_j$ respectively, $\tau_i$ is the ratio of ion to electron temperature).

Table 1. Basic equilibrium parameters

| $a$ [m] | $R$ [m] | $r_j$ [cm] | $L_n$ [cm] | $L_{T_e}$ [cm] | $B$ [T] |
|---|---|---|---|---|---|
| 0.6 | 1.7 | 54 | 12 | 6 | 1.8 |
| $n_e$ [$10^{19}$m$^{-3}$] | $T_e$ [eV] | $q$ | $\hat{s}$ | $n$ | $\tau_i$ | $\delta_e$ |
| 1.2 | 140 | 3 | 1.5 | -8 | 0.5 | 0.1 |

At large $k$, the ballooning equation equation (21) becomes a Weber-Hermite equation allowing the asymptotic boundary condition for an outgoing wave [16]

$$\chi(k,\lambda) \to \exp\left[i\sqrt{\frac{(1+\bar{\eta}_i/\hat{\omega})}{(1+\delta_\tau)}}\frac{\hat{\omega}k_g\hat{s}}{\hat{\omega}_s}\frac{k^2}{2}\right]. \quad (27)$$

The two coupled 1D ordinary differential equation (21) and equation (23) will be solved by an iterative method, because the ballooning operator $L_0(k,\lambda;\hat{\omega})$, and the averaged $\bar{L}_1(k;\hat{\omega})$ and $\bar{L}_2(k;\hat{\omega})$ all contain the (2D global) eigenvalue $\hat{\omega}$ which remains unknown until the equation (23) for $\psi(\lambda)$ is solved. The steps in the iterative procedure (same as in [15]) are listed below:

(I) Begin with an initial guess, $\hat{\omega} \to \hat{\omega}^{(0)}$ to solve equation (21),

$$\left[L_0\left(k,\lambda;\hat{\omega}^{(0)}\right) - \Omega^{(0)}(\lambda)\right]\chi^{(0)}(k,\lambda) = 0 \quad (28)$$

by imposing the boundary condition equation (27) for all $\lambda$.

(II) Substitute $\chi^{(0)}(k,\lambda)$ into equation (26) to compute $\bar{L}_1^{(0)}(\lambda;\hat{\omega}^{(0)})$ and $\bar{L}_2^{(0)}(\lambda;\hat{\omega}^{(0)})$, and consequently obtain the equations for $\Phi^{(0)}(\lambda)$ (see Appendix B)

$$\frac{d^2\Phi^{(0)}(\lambda)}{d\lambda^2} + \frac{n^2}{\bar{L}_2^{(0)}(\lambda;\hat{\omega}^{(0)})}\left\{\Omega^{(0)}(\lambda) - \Omega\left(\hat{\omega}^{(1)}\right) + \frac{\left[\bar{L}_1^{(0)}(\lambda;\hat{\omega}^{(0)})\right]^2}{4\bar{L}_2^{(0)}(\lambda;\hat{\omega}^{(0)})}\right\}\Phi^{(0)}(\lambda) = 0. \quad (29)$$

This equation is solved with evanescent boundary conditions.

(III) The global eigenvalue $\hat{\omega}^{(1)}$ follows from equation (17) by substituting the eigenvalue of equation (29) $\Omega(\hat{\omega}^{(1)})$.

Repeat the procedures (I-III) to obtain $\hat{\omega}^{(i+1)}$ from $\hat{\omega}^{(i)}$ until $\left|1 - \hat{\omega}^{(i+1)}/\hat{\omega}^{(i)}\right| < \varepsilon$ with $\varepsilon \equiv 10^{-4}$ as the convergence condition.

The convergence was usually achieved after 5 iterations. The 2D eigenvalue $\hat{\omega}_{\text{WABT}} = 0.758 + 0.068i$. It is close to the 1D ballooning solution at $\lambda = 0$, $\hat{\omega} = 0.793 + 0.073i$ (in both the real frequency and the growth rate). The potential structure $V(k,\lambda)$ and wave function $\chi(k,\lambda)$ at $\lambda = 0$ and $\lambda = \pm 0.74$ are displayed in figure 1. Here the potential structure is defined in equation (18): $L_0(k,\lambda;\hat{\omega}) \to \partial^2/\partial k^2 - V(k,\lambda)$.

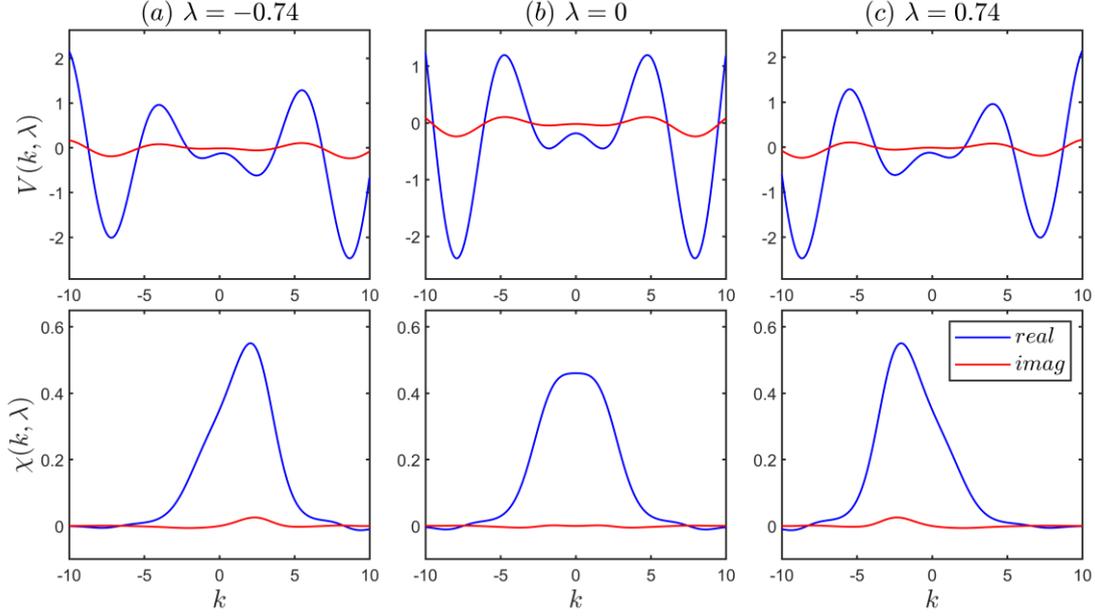

Figure 1. Potential structure $V(k,\lambda)$ and wave function $\chi(k,\lambda)$ at

(a) $\lambda = -0.74$, (b) $\lambda = 0$ and (c) $\lambda = 0.74$

In figure 2 the averaged $\bar{L}_1^{(0)}(\lambda)$, $\bar{L}_2^{(0)}(\lambda)$, and $\Omega(\lambda)$ are presented, the blue and red lines stand, respectively, for the real and imaginary parts. One can see that, $\bar{L}_1^{(0)}(\lambda)$ is almost a constant, $\bar{L}_2^{(0)}(\lambda)$ varies in $\lambda$, but only slightly. The second small parameter of WABT $\Xi \equiv \bar{L}_1/2\bar{L}_2 \approx 0.06$ is small enough to warrant the validity of WABT (see the end of Appendix B). The potential $\Omega(\lambda)$ is an even function of $\lambda$.

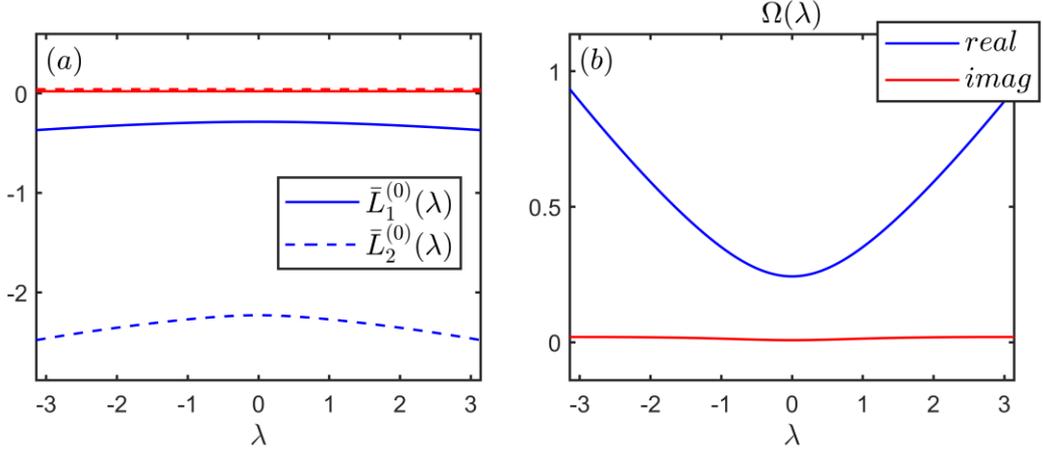

Figure 2. (a) average higher order coefficient $\bar{L}_1^{(0)}(\lambda)$, $\bar{L}_2^{(0)}(\lambda)$, (b) local eigenvalue $\Omega(\lambda)$

The Floquet phase $\psi(\lambda)$ is shown in figure 3. The real part of $\psi(\lambda)$ is a Gaussian located at $\lambda = 0$. The imaginary part looks like a dipole with two peaks around $\lambda = \pm 0.74$ and breaks the $\lambda$-inversion symmetry (resulting from the small parameter $\Xi$).

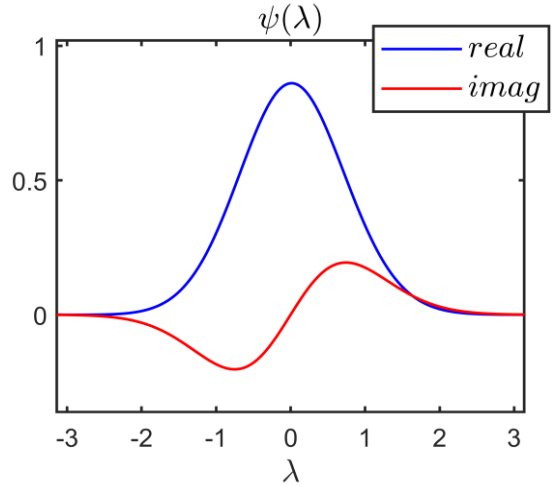

Figure 3. the Floquet phase distribution (FPD) $\psi(\lambda)$

The two small parameters required by the WABT structure (defined at the end of Appendix B) are, *a posteriori*, shown to be small: $\sqrt{\Omega_1 / 2n\bar{L}_2} \approx 0.08 \ll 1$ and $\Xi\sqrt{n} \approx 0.18 \ll \pi/2$ ($\bar{L}_1 \approx -0.3$, $\bar{L}_2 \approx -2.2$ and $\Omega_1 \approx -0.25$).

## 4. Iterative finite difference method for $\delta_e$ mode

In order to expose the spatial structure of the $\delta_e$ mode, we must convert the 2D wave function $\varphi(k,\lambda) = \chi(k,\lambda)\psi(\lambda)$ (obtained in ballooning space) into the $(x,l)$ representation by making use of the 2D Fourier-ballooning transform equation (15),

$$\bar{\chi}(x-l,\lambda) = \int_{-\infty}^{+\infty} dk \exp[ik(x-l)]\chi(k,\lambda), \tag{30}$$

and

$$\varphi_l(x) = \frac{1}{2\pi}\int_{-\pi}^{\pi} d\lambda \exp(-il\lambda)\psi(\lambda)\bar{\chi}(x-l,\lambda). \tag{31}$$

On substituting $\bar{\chi}(x-l,\lambda)$ into equation (31) and integrating over $\lambda$, we get the WABT wave functions $\varphi_l(x)$ corresponding to the poloidal mode number $l$; $\varphi_l(x)$ for various $l = -4,-3,...,4,5$ are displayed in figure 4(a). This solution can be used as an initial guess in the shifted inverse power method [26] to solve the 2D eigenvalue problem equation (14) on the $(x,l)$ grid by the iterative finite difference method. In fact, the WABT solution provides not only the radial boundary of each rational surface to be outgoing waves, but also the phase relation between neighboring rational surfaces, as natural boundary condition for the 2D local eigenmode [27].

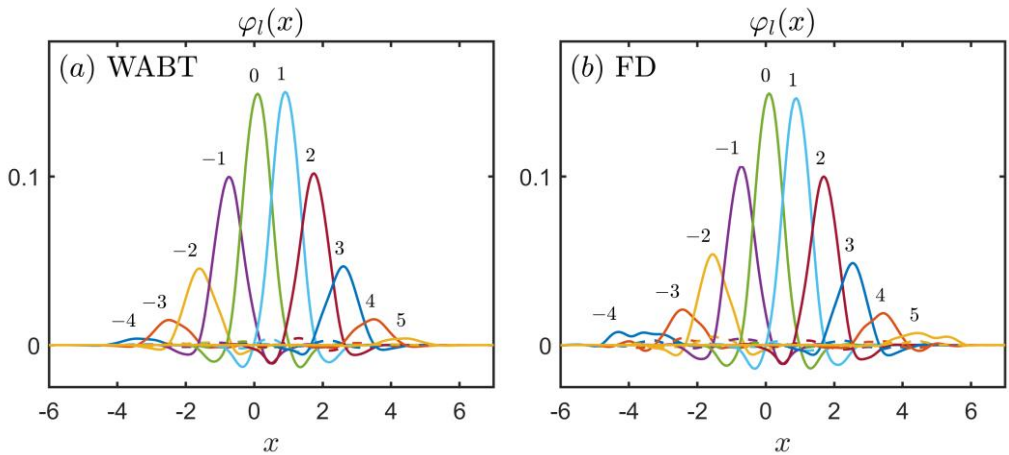

Figure 4. wave functions $\varphi_l(x)$ ($l = -4, -3, ..., 4, 5$) of (a) WABT and (b) iterative finite difference solution, the solid and dashed lines are real and imaginary part respectively

The 2D eigenvalue problem equation (14) is put in the form as $T[\partial/\partial x, l; \hat{\omega}]\varphi_l(x) = \Omega(\hat{\omega})\varphi_l(x)$, where $T$ is a differential operator with derivative of $x$, it also depends on $l$ and the eigenvalue $\hat{\omega}$. The spatial discrete grids are $x_k \equiv k \cdot h$ ($k = -K, -K+1, ..., K-1, K$), the step size is $h = (x_r - x_l)/2K$, $x_l$ ($x_r$) is the left (right) boundary. The $l$ grids are $l = -L, -L+1, ..., L-1, L$, and $\varphi_l(x)$ is cut-off at large $|l| > L$. Use is made of the central difference for derivative of $x$ to yield the matrix equation

$$\mathbf{M}(\hat{\omega}) \cdot \mathbf{\Phi} = \Omega(\hat{\omega})\mathbf{\Phi}, \tag{32}$$

where $\mathbf{\Phi} = (\mathbf{\Phi}_{-L}, \mathbf{\Phi}_{-L+1}, \cdots, \mathbf{\Phi}_{L-1}, \mathbf{\Phi}_L)^{\mathrm{T}}$, $\mathbf{\Phi}_l = (\varphi_{l,-K}, \varphi_{l,-K+1}, \cdots, \varphi_{l,K-1}, \varphi_{l,K})^{\mathrm{T}}$, $\varphi_{l,k} \equiv \varphi_l(x_k)$. $\mathbf{M}$ is a block tri-diagonal matrix and its dimension is $(2L+1)(2K+1) \times (2L+1)(2K+1)$.

The eigenvalue problem equation (32) can only be solved by an iterative method because it is nonlinear in $\hat{\omega}$. The iterative procedure is listed below:

(I) It begins with the eigenvalue of WABT solution as initial guess, $\hat{\omega}_{\text{WABT}} \to \hat{\omega}^{(0)}$ to compute the matrix $\mathbf{M}^{(0)} = \mathbf{M}(\omega^{(0)})$ and eigenvalue $\Omega^{(0)} = \Omega(\omega^{(0)})$ (see equation (17)).

(II) Wave functions $\varphi_l(x)$ of WABT solution at discrete grids $\mathbf{\Phi}^{(0)}$ are normalized, $\mathbf{\Phi}^{(0)} = \mathbf{\Phi}^{(0)}/\|\mathbf{\Phi}^{(0)}\|_2$, the matrix equation $(\mathbf{M}^{(0)} - \Omega^{(0)}) \cdot \mathbf{\Phi}^{(1)} = \mathbf{\Phi}^{(0)}$ is solved for $\mathbf{\Phi}^{(1)}$ by LU decomposition.

(III) Compute $\Delta\Omega^{(1)} = (\mathbf{\Phi}^{(1)})^{\dagger} \cdot (\mathbf{M}^{(0)} - \Omega^{(0)}) \cdot \mathbf{\Phi}^{(1)} / (\mathbf{\Phi}^{(1)})^{\dagger} \cdot \mathbf{\Phi}^{(1)}$ and go back to step (II), $\mathbf{\Phi}^{(0)} \to \mathbf{\Phi}^{(1)}$ until $\delta = \|\Delta\Omega^{(1)} \cdot \mathbf{\Phi}^{(1)} - \mathbf{\Phi}^{(0)}\|_{\infty} < 10^{-6}$.

(IV) Substitute the new eigenvalue $\Omega^{(1)} = \Omega^{(0)} + \Delta\Omega^{(1)}$ into equation (17) to obtain the global eigenvalue $\hat{\omega}^{(1)}$.

In the preceding lines, $\boldsymbol{\Phi}^{\dagger}$ stands for the Hermitian conjugate of $\boldsymbol{\Phi}$, $\|\boldsymbol{\Phi}\|_2 \equiv \sqrt{\sum_i \Phi_i^2}$ and $\|\boldsymbol{\Phi}\|_\infty \equiv \max_i |\Phi_i|$ are the 2-norm and $\infty$-norm of vector $\boldsymbol{\Phi}$ respectively.

(V) Repeat the steps (I-IV) to obtain $\hat{\omega}^{(i+1)}$ from $\hat{\omega}^{(i)}$ until $\left|1 - \hat{\omega}^{(i+1)}/\hat{\omega}^{(i)}\right| < \varepsilon$, with $\varepsilon \equiv 10^{-4}$ as the convergence condition.

After 5 iterations, the eigenvalue converged to $\hat{\omega}_{\text{FD}} = 0.745 + 0.065i$. The difference of iterative finite difference solution and WABT solution, $\left|\hat{\omega}_{\text{FD}} - \hat{\omega}_{\text{WABT}}\right|/\left|\hat{\omega}_{\text{WABT}}\right| \approx 1.7\%$, was within the expected error ($\sim 1/n = 12.5\%$). The wave functions $\varphi_l(x)$ of the iterative finite difference solution, displayed in figure 4(b), are in good agreement with the WABT solution.

The wave function in $(r, \vartheta)$ representation, $\varphi(r, \vartheta) = \exp(-im\vartheta)\sum_l \varphi_l(x)\exp(-il\vartheta)$, is shown in figure 5; the 2D mode structure $\varphi(r, \vartheta)$ and the close-up in bad curvature region for both WABT and iterative finite difference solution are highlighted. The mode structure, embodied in the wave function, is the crucial input to compute Reynolds stress and group velocity of $\delta_e$ mode in section 5.

One may have noticed that the dumbbell radial structure shown in figure 5 is consistent with the almost equal height of $l = 0$ and $l = 1$ in figure 4, feathering inward shift of the EDW mode structure. The position of $r_j = 54\text{cm}$ is equivalent to $\rho = 0.9$. The actual 'center of mode' is always less than $\rho = 0.9$ as shown in figure 5, implying an inward shift. It is consistent with the results shown in figure 4. Positive sideband number corresponds to radial shift inward in our convention that the toroidal number is negative. For example, center of mode for $l = 1$ in figure 4 is around $x = 1$, corresponding to radial shift $\Delta\rho = -0.025$.

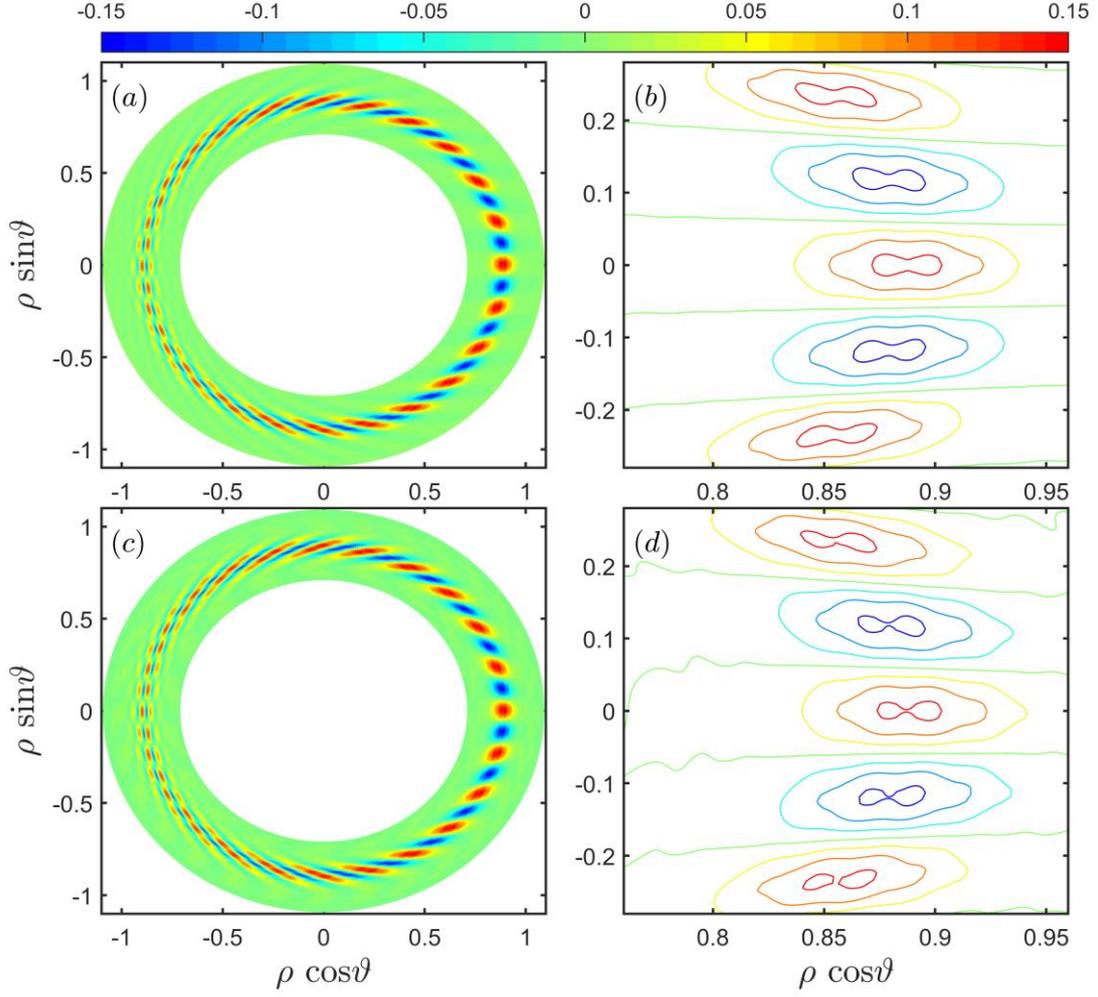

Figure 5. 2D mode structure $\varphi(r,\vartheta)$ of (a) WABT solution, (b) its close-up in bad curvature regime, and (c) iterative finite difference solution, (d) its close-up in bad curvature regime, where $\rho \equiv r/a$ is the normalized minor radius

## 5. Reynolds stress and group velocity for $\delta_e$ - mode

The Reynolds stress is defined by $\mathfrak{R}_\vartheta \equiv \langle \tilde{u}_r \tilde{u}_\vartheta \rangle$, where $\langle \ldots \rangle$ stands for ensemble average, which is equivalent to the average over the poloidal angle [27,28]. For perturbed electrostatic potential $\tilde{\boldsymbol{u}} = \rho_s c_s \boldsymbol{b} \times \nabla \varphi$, the explicit expression of (poloidal) Reynolds stress pertaining to the rational surface $r_j$ is

$$\Re_{\vartheta,j}(r) = -\rho_s^2 c_s^2 \oint d\vartheta \frac{\partial \varphi(r,\vartheta)}{r_j \partial \vartheta} \frac{\partial \varphi(r,\vartheta)}{\partial r}. \tag{33}$$

The intensity of drift wave for $\delta_e$-model is defined as $I_m(r) \equiv \oint d\vartheta \varphi(r,\vartheta) \varphi(r,\vartheta)$, here $I_m(r)$ describes the radial distribution of drift wave. When computing the physical quantity, we only retain the real part of $\varphi(r,\vartheta)$.

The Reynolds stress and intensity of drift wave, computed by the 2D mode structure of WABT and iterative finite difference solution in section 4 by direct numerical integration, are displayed in figure 6, where $I(r) \equiv I_m(r)/I_m(r_m)$ and $R(r) \equiv \Re_{\vartheta,j}(r)/\rho_s^2 c_s^2 k_\vartheta^2 \hat{s} I_m(r_m)$, $r_m$ is the peak position of $I_m(r)$. One can see that $I(r)$ is a Gaussian envelop plus fine-scale radial structure (due to sideband coupling). The Reynolds stress $R(r)$ has the shape of monopole instead of dipole obtained earlier in the fluid ITG model [27].

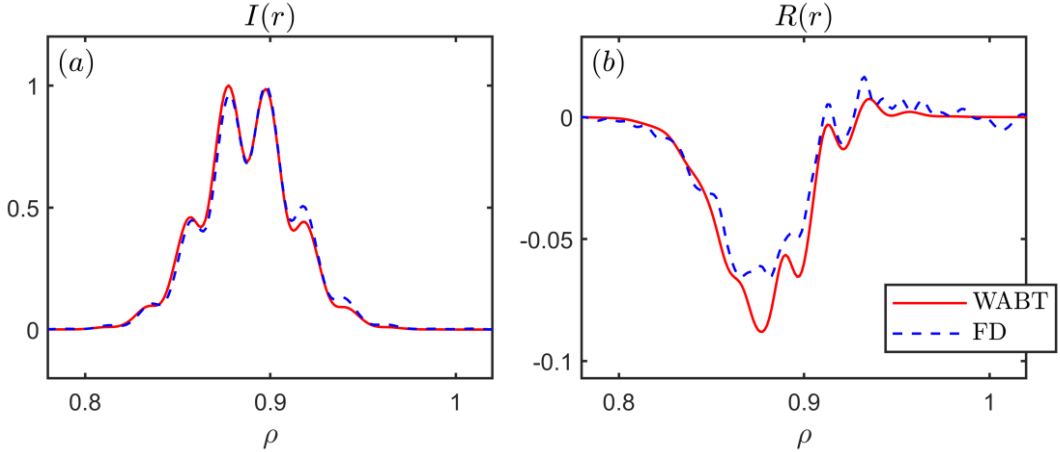

Figure 6. (a) the intensity of drift wave, (b) Reynolds stress computed by WABT solution (red solid lines) and iterative finite difference solution (blue dashed lines)

The radial and poloidal group velocities are functions only of the poloidal angle because the drift wave in tokamak is a standing wave in radial direction. The characteristic line in poloidal direction of equation (10) is given by $d\vartheta/dt = v_{gy}(\vartheta)/r_j$, that yields mapping time to poloidal angle $\vartheta(t)$. As a result, the radial group velocity becomes a periodic function of time, $\upsilon_{gr}(t)=\upsilon_{gr}(\vartheta(t))$. To explore the structure of group velocity, the average of three operators over fast scale shown in equations (11), (12) must be computed:

$$\langle K_r \rangle \equiv -ik_\vartheta \hat{s} \frac{\int dx \varphi^*(x,\vartheta) \partial_x \varphi(x,\vartheta)}{\int dx \varphi^*(x,\vartheta) \varphi(x,\vartheta)}, \quad (34)$$

$$\langle K_r^2 \rangle \equiv -k_\vartheta^2 \hat{s}^2 \frac{\int dx \varphi^*(x,\vartheta) \partial_x^2 \varphi(x,\vartheta)}{\int dx \varphi^*(x,\vartheta) \varphi(x,\vartheta)}, \quad (35)$$

$$\langle \nabla_\parallel^2 \rangle \equiv -\frac{1}{q^2 R^2} \frac{\int dx \varphi^*(x,\vartheta) \sum_l (x-l)^2 \varphi_l(x) \exp(-il\vartheta)}{\int dx \varphi^*(x,\vartheta) \varphi(x,\vartheta)}. \quad (36)$$

It is done by substituting the 2D mode structure of WABT solution and iterative finite difference solution into equations (34-36); the three averaged normalized operators $\hat{K}_r \equiv \langle K_r \rangle / k_\vartheta$, $\hat{K}_r^2 \equiv \langle K_r^2 \rangle / k_\vartheta^2$ and $\hat{K}_\parallel^2 \equiv \langle \nabla_\parallel^2 \rangle / k_\vartheta^2$ are displayed in figure 7.

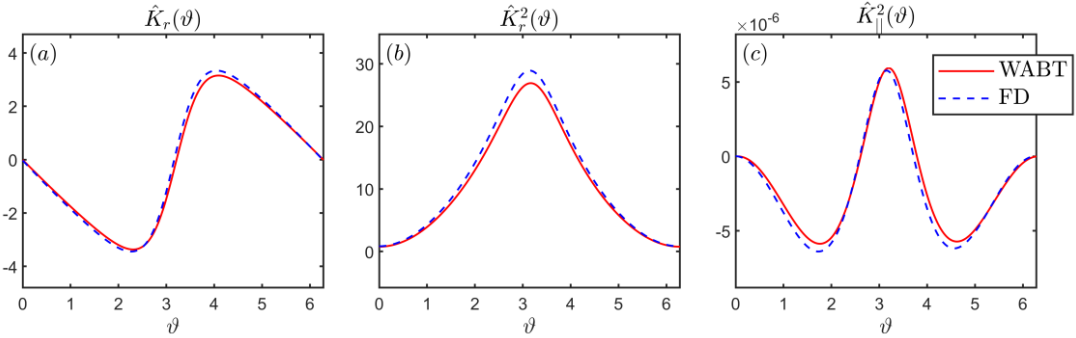

Figure 7. Three normalized operators, (a) $\hat{K}_r \equiv \langle K_r \rangle / k_\vartheta$, (b) $\hat{K}_r^2 \equiv \langle K_r^2 \rangle / k_\vartheta^2$, (c) $\hat{K}_\parallel^2 \equiv \langle \nabla_\parallel^2 \rangle / k_\vartheta^2$, the red solid and blue dashed line are solution of WABT and iterative finite difference respectively

For real and regular group velocities in stationary background micro-turbulence, the drift wave envelope $|\bar{\phi}|$ is invariant [8], it is therefore convenient to take the eikonal form $\bar{\phi} \sim \exp(i\Theta)$ with real $\Theta$. The imaginary part $i\delta_e$ and the growth rate ($\mathrm{Im}\,\omega$) in equations (11-12) won't contribute to group velocity. In figure 8, the $\vartheta$ dependence of the radial and poloidal group velocity is displayed. Use is made by $t = \int_0^{\vartheta} r_j d\vartheta' / \upsilon_{gy}(\vartheta')$ to generate the mapping $\vartheta(t)$, then the radial group velocity is plotted as function of time $t$ as shown in figure 8(c). One can see that the period of radial group velocity is about 7ms. Each period consists of two consecutive distinct phases: a long slowly varying and a short rapid zero-crossing. There is a little difference from the counterpart of ITG [8]. For ITG group velocity, both up-zero-crossing and down-zero-crossing are rapid. For $\delta_e$-mode, the rapidity of up-zero-crossing is high, but is much lower for down-zero-crossing.

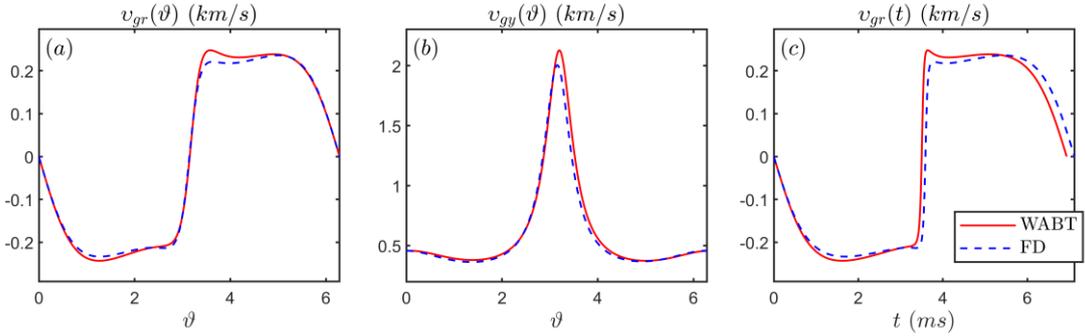

Figure 8. (a) radial, (b) poloidal group velocities versus $\vartheta$, (c) radial group velocity versus $t$

Now, we are ready to study the interplay between zonal flow and phase function in meso-scale provided that the ambient turbulence is EDW. The study is to solve the zonal flow equation and the envelope equation jointly by making use of the Reynolds stress and group velocity obtained in this section for EDW.

## 6. Remarks on Instanton in $\delta_e$-mode

In a torus like tokamak the geodesic curvature introduces high frequency branch into the zonal flow equation via coupling to sound wave. It also modifies the slab low frequency zonal flow (SLFZF) into TLFZF. This has been shown in literature for electron drift wave [29] (also see equation (C10) in Appendix C). However, the Reynolds stress as the source of zonal flow equation is also modulated by the drift wave envelope, not a static one. It is precisely the purpose of this section to demonstrate that such a modulation plays the key role to intermittent excitation of GAM. The modulation by drift wave envelope is to replace $\mathfrak{R}_{\vartheta,j}(r)$ by $\mathfrak{R}_{\vartheta,j}(r)\cos^2\Theta$ in the zonal flow equation, where $\Theta$ consists of two phases [8]. A long-lived standing wave phase, which called the Caviton and a short-lived traveling wave phase (in radial direction), called the Instanton. In Caviton phase $\Theta$ varies slowly in time. The corresponding modulation has little impact on the high frequency branch of zonal flow (GAM). Right after the group velocity crosses zero, the Caviton transits to Instanton as a linear traveling wave in radial. Then, it evolves to nonlinear stage rapidly with increasing frequency. As soon as the frequency is resonant to GAM frequency, the GAM is excited. The key processes leading to GAM excitation will be presented in this section.

When the modulation is included in the TLFZF equation in collisional regime given by [30, 31] and equation (C10) in Appendix C, it becomes

$$\left(1+2q^2\right)\frac{\partial \overline{\upsilon}}{\partial t} - \mu \frac{\partial^2 \overline{\upsilon}}{\partial r^2} + \frac{\partial}{\partial r}\left[\mathfrak{R}_{\vartheta,j}(r)\cos^2\Theta\right] = 0, \tag{37}$$

where $\Theta$ represents the phase (eikonal) of drift wave envelope $\overline{\phi} = \exp(i\Theta)$ in equation (10), and $\mathfrak{R}_{\vartheta,j}(r)$ is the Reynolds stress defined in equation (33). $\mu$ is the perpendicular viscosity; for classical fluid model, $\mu \to \mu_B \equiv 3\nu_{ii}\rho_i^2/10$, where $\nu_{ii}$ is the ion-ion collision frequency and $\rho_i$ is the ion Larmor radius. For a real plasma, the viscosity is likely to be anomalous, $\mu = a_\mu \mu_B$, $a_\mu$ being the measure of anomaly, which is set to be $a_\mu = 3$.

As derived in [8], a formal inhomogeneous solution of equation (10) is

$$\Theta(r,t) = k_g \int_0^t dt' \overline{\upsilon}\left(r - \int_{t'}^t ds \upsilon_{gr}(s), t'\right), \tag{38}$$

where $\upsilon_{gr}(s)$ is the radial group velocity of EDW as a function of time, which has been calculated and shown in figure 8(c). In this section the numerical results are obtained from the solution of equations (37-38). The numerical results related to GAM excitation will be presented in the next section 7. The normalization and numerical methods for the zonal flow equation set are introduced in Appendix D and will not be repeated elsewhere.

## 6.1 Correlation between occurrence of spikes and zero-crossing of radial group velocity

Equations (37-38) are numerically solved with parameters listed in Table 1, and $I_m(r_m)$, the intensity of turbulence, is chosen to be $10^{-3}$.

The vertical dotted line 1 (2) in figures 9,10 and figures 14-17 denotes the time point upon downward (upward) zero-crossing of radial group velocity. One can clearly see the huge differences in temporal structure between near zero-crossing and away from zero-crossing. Near zero-crossing structure looks spiky in $\Theta$, kinky in $\overline{\upsilon}$; away from zero-crossing structure looks smooth in both $\Theta$ and $\overline{\upsilon}$ as shown in figure 9(b) and 9(c). The similar structures have been shown in figure 6 of [8] for ambient turbulence of ITG, where the spiky phase is named Instanton and the smooth phase is named Caviton. The nomenclature can be understood by seeing the spatial structure in figure 10 of [8]. The close-up temporal structure of $\Theta$ for 5 radial positions is displayed in figure 10(a) and 10(b), very much like that shown in figure 7 of [8]. By picking up three radial positions "yellow ($\rho = 0.90$) – red ($\rho = 0.89$) – blue $(\rho = 0.88)$" in figure 10(a) one can see a traveling wave moving inward. By the same token, what shown in figure 10(b) is a traveling wave moving outward. The 3D structures shown in figure 10(c) and 10(d) depict spatiotemporal structure of EDW envelope square $\cos^2 \Theta$ in real space. Although the intensity represented by color does not seem accurate to satisfaction (dark region means $\Theta = 0$), it does have roughly shown the locus of the traveling wave; moving inward is shown in figure 10(c) and moving outward is shown in figure 10(d). Henceforth, moving inward (outward) traveling wave is referred to be ingoing (outgoing) Instanton.

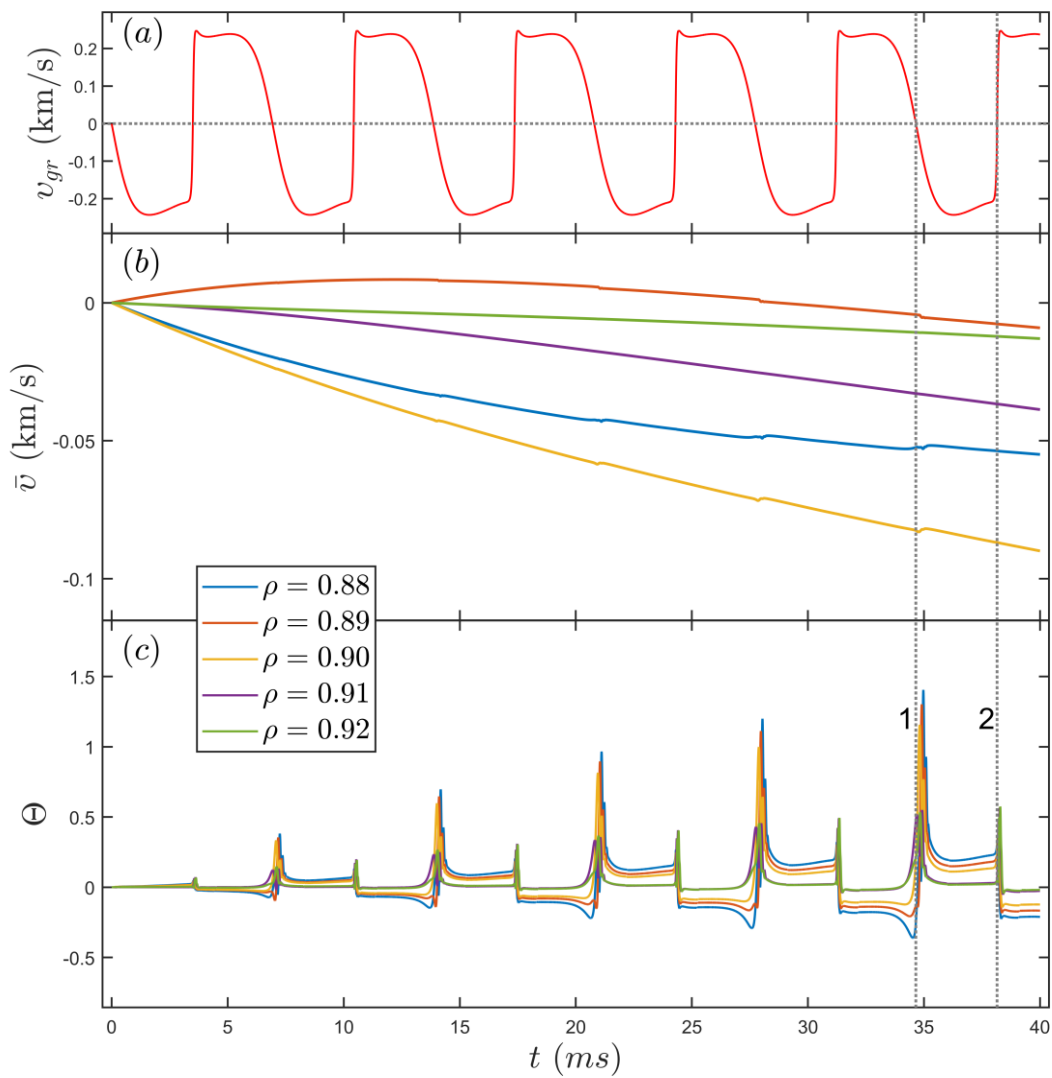

Figure 9. The time evolution of 40 ms for 5 distinct spatial positions to display the relationship of zonal flow kinks, phase spikes and zero-crossing of radial group velocity for $I_m(r_m) = 10^{-3}$

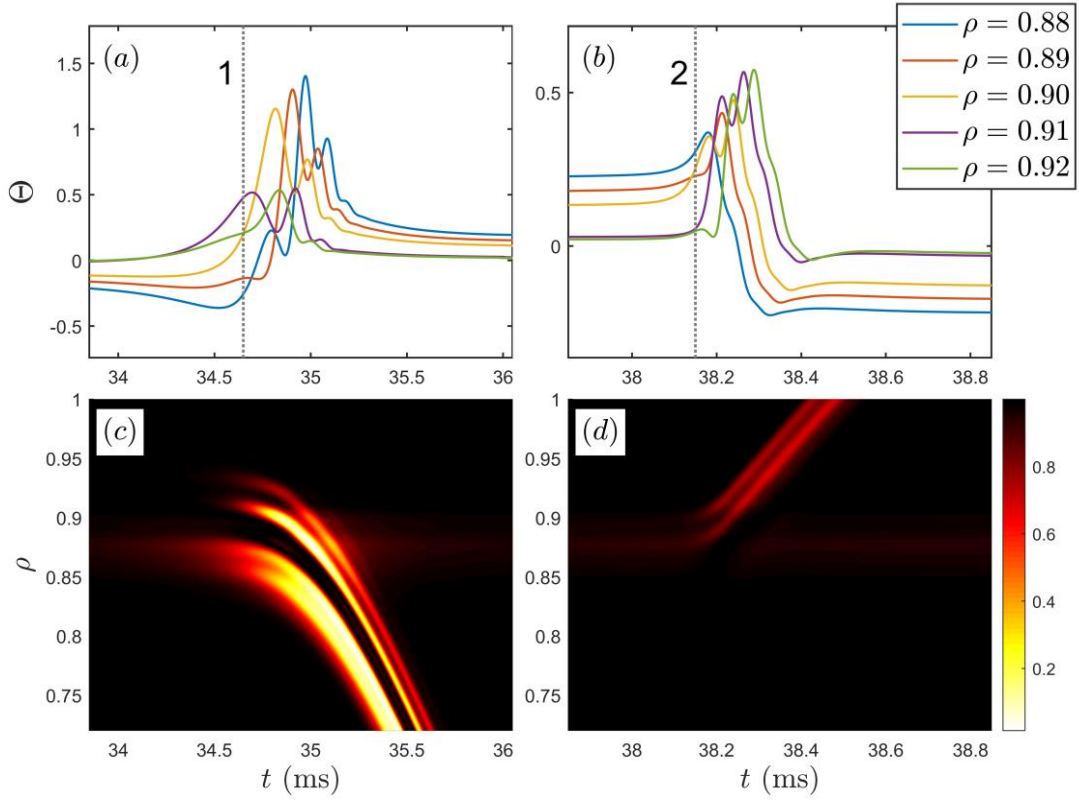

Figure 10. The time evolution of phase $\Theta$ and spatiotemporal structure of EDW envelope square $\cos^2\Theta$ near period 1 and 2 in figure 9

## 6.2 The transition of Caviton to Instanton

The Caviton refers to the phase of drift wave envelope as a slowly moving nonlinear standing wave (smooth phases between spikes in figure 9(c)). The spatial structure of Caviton is shown in figure 11 at three moments 33ms, 37ms and 40ms. They look like monopole structure in contrast to the dipole structure for ITG ambient turbulence (figure 4 of [8]). The three moments in time are chosen to accommodate two transitions – at 34.65ms, and at 38.15ms. The cyan line marks the structure before transition 1. The blue line marks the structure between transition 1 and 2. The red line marks the structure after transition 2. Apparently, transition 1 alters the negative-Caviton (negative peak) to positive-Caviton (positive peak) and transition 2 alters the positive-Caviton back to negative-Caviton without obvious change in shape. Such a change of spatial structure can also be seen in figure 10(a) from 34ms to 35.5ms, *e.g.*, the bottom blue line at 34ms becomes the top blue line at 35.5ms, similarly in figure 10(b) from 38ms to 38.6ms. As a rule of thumb the relationship between the sign of the Caviton and zero-crossing is that the downward (upward) zero-crossing leads to positive (negative) sign of Caviton. This rule is valid as long as the sign of radial group velocity remains the same until next zero-crossing occurs (in this paper $k_g < 0$). The spatial structure of Caviton during the period of time when away from the zero-crossing, is approximately determined by radial integration over the zonal flow ($\upsilon_{gr}(t)\left[\partial\Theta(r,t)/\partial r\right]=k_g\bar{\upsilon}(r,t)$), because the time variation of Caviton is small in the envelope equation and can be neglected in that time duration, as shown in figure 10(a) and (b). As a result, the dipole like zonal flow leads to monopole Caviton.

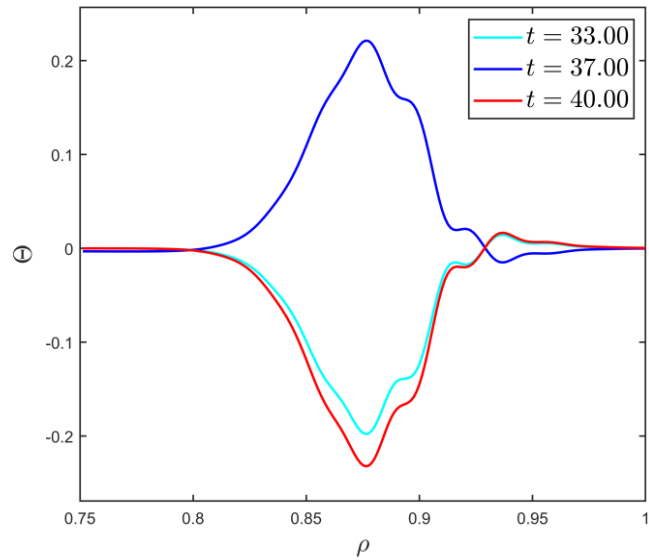

Figure 11. The spatial structure of Caviton at 33ms, 37ms and 40ms

In figure 12 displayed is the spatiotemporal structure of $\Theta$ around transition 1 and 2, where red-black and blue-white represent crest and trough respectively. The travelling waves of ingoing and outgoing Instanton are marked clearly by the strokes in non-green colors. The alternative sign of Caviton is displayed by the change in color within the reaction region, corresponding to $r \in [50,54]$ cm in figure 12. The change from yellow to cyan (cyan to yellow) represents the change of positive to negative (negative to positive) Caviton.

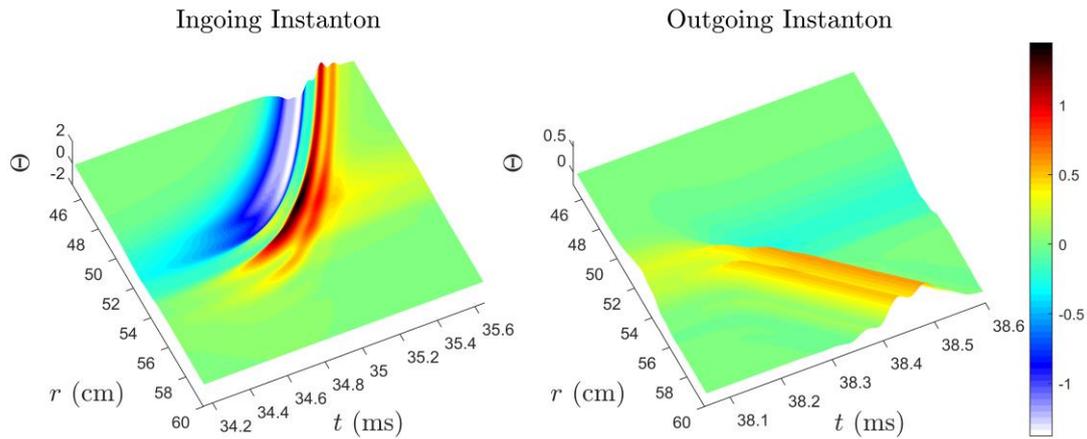

Figure 12. The spatiotemporal structure of $\Theta$ for EDW

The transition of Caviton to Instanton can also be presented in the movie 'spatiotemporal evolution' in the caption of figure 13 (Multimedia view). This presentation demonstrates vividly that Caviton is a slow nonlinear standing wave and Instanton is a fast traveling wave. Six snapshots of the movie are selected as shown in figure 13. In figure 13(a) displayed is the Caviton starts to grow at 34.20ms from $\Theta \approx 0$. The growing continues until 34.65ms when $\upsilon_{gr}$ crosses zero. The early stage Instanton is shown in figure 13(b). Then, the ingoing Instanton is shown in figure 13(c). One can clearly see that the major portion of Instanton moves outside the reaction region till it totally disappears. In figure 13(d-e) displayed are the similar process, however, for outgoing Instanton. One may have noticed that the intensity of outgoing Instanton is much weaker than the outgoing one. This is consistent with the intensity shown in figure 10(c) and (d). It could be attributed to the fact that the lifetime of ingoing (outgoing) Instanton is about 0.55ms (0.25ms) due to lower (higher) rapidity of zero-crossing.

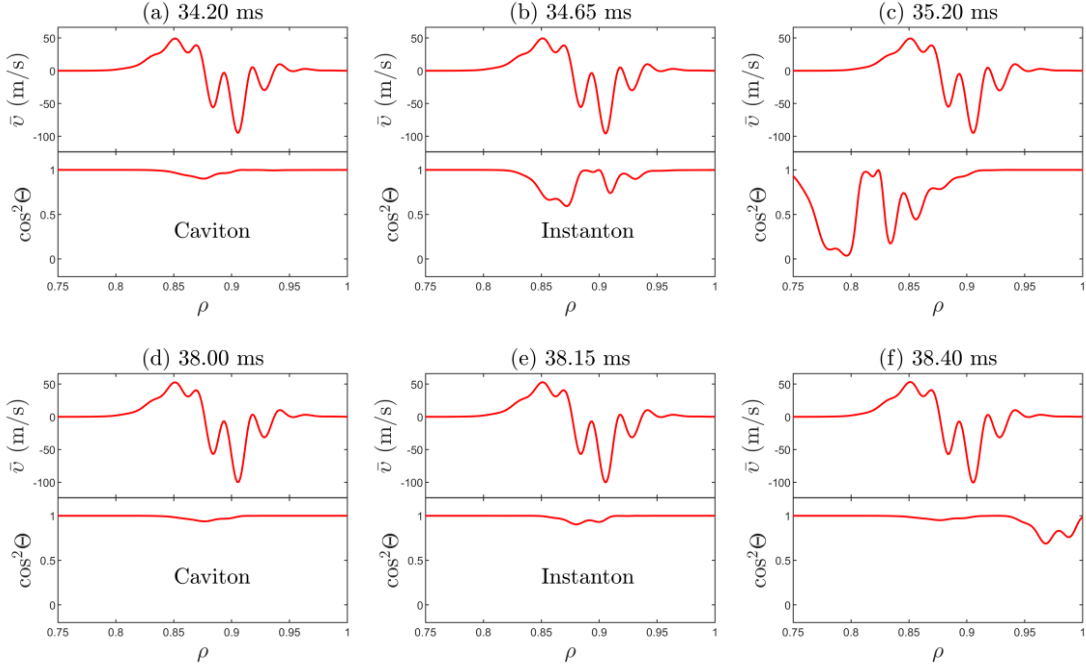

Figure 13. Snapshots of zonal flow $\bar{\upsilon}$ and EDW envelope square $\cos^2\Theta$ of 6 time points. The time evolution for 34-40ms can be seen via the link[1] 'spatiotemporal evolution' (Multimedia view)

Now the general information regarding transition from Caviton to Instanton can be summarized as follows. Before the downward zero-crossing 1 occurs, the eikonal $\Theta$ is a negative Caviton in shape of monopole. The zero-crossing 1 induces transition from Caviton to ingoing Instanton. As it disappears, a new positive Caviton emerges to grow. It lasts till upward zero-crossing 2 occurs. The zero-crossing 2 induces transition from positive Caviton to outgoing Instanton. As it disappears in reaction region, a new negative Caviton emerges to grow.

Before ending this sub-section, it may be appropriate to briefly describe the physics involving transition of Caviton to Instanton, and the key role played by zero-crossing of radial group velocity.

---

[1] http://staff.ustc.edu.cn/~lzy0928/spatiotemporal%20evolution.mp4

As shown in equation (38), the zonal flow $\bar{\upsilon}$ modulates drift wave envelope in phase ($\bar{\phi} \sim \cos\Theta$, where $\bar{\phi}$ is the drift wave envelope). The radial group velocity $\upsilon_{gr}$ appears as the argument of $\bar{\upsilon}$ under integral, describing the movement of drift wave envelope along the radial characteristic line. According to equation (38) the finite $\upsilon_{gr}$ makes a shift of radial argument of $\bar{\upsilon}$ from $r$ to $r - \int_{t'}^{t} ds \upsilon_{gr}(s)$. As shown in figure 9(a) the $\upsilon_{gr}$ always consists of two consecutive distinct phases: a long slowly varying phase having a large $|\upsilon_{gr}|$ on top and button and a short phase with rapid zero-crossing. Notice that the zonal flow $\bar{\upsilon}$ is localized around the region where the Reynolds stress is not small (reaction region). Before crossing zero, $\upsilon_{gr}$ is large, the shift would make the argument of $\bar{\upsilon}$ run out of the reaction region depending on the reference position $r$, where $\bar{\upsilon}$ is too small to contribute to the integral in equation (38). This process corresponds to formation of Caviton as $\Theta$ is slowly varying. Upon $\upsilon_{gr}$ zero-crossing, the sign is changed, making the shift smaller, and pulling the local integrand $\bar{\upsilon}$ back to the reaction region, making $\bar{\upsilon}$ contribute to $\Theta$ again. Such a process occurs on different instants at different reference positions; it simulates wave propagation, just like pattern propagation shown in figure 10(a), (b).

### 6.3 The nonlinear traveling Instanton with increasing frequency to decades kHz

In this sub-section we begin with showing the analytic solution of Instanton near zero-crossing, and compare it with the numerical solution. This study helps us to understand that the spatial structure of zonal flow plays key role in the nonlinear evolution of Instanton, even though it does not change very much during the lifetime of Instanton. We then shall demonstrate that the nonlinear Instanton may evolve to the state containing the spectrum to decades kHz range.

Equation (38) can be cast into the following form

$$\Theta(r,t) = \Theta(r,t_g) + k_\vartheta \int_{t_g}^{t} dt' \bar{\upsilon}\left(r - \int_{t'}^{t} ds \upsilon_{gr}(s), t'\right). \tag{39}$$

Denoting $t_g$ to be the instance at zero-crossing of radial group velocity $\upsilon_{gr}(t_g) = 0$, and assuming the interval $[t_g, t]$ is sufficiently small $\approx \varepsilon$, Then $\upsilon_{gr}(s)$ is expanded up to linear function

$$\upsilon_{gr}(s) = \left(\frac{\partial \upsilon_{gr}}{\partial s}\right)_{t_g} (s - t_g) + \ldots . \tag{40}$$

Defining $r_g(t,t',t_g) \equiv \int_{t'}^{t} ds \upsilon_{gr}(s)$, and making use of equation (40) to obtain

$$r_g(t,t',t_g) = \frac{1}{2}\left(\frac{\partial \upsilon_{gr}}{\partial s}\right)_{t_g} \left[(t+t') - 2t_g\right](t-t') \sim \varepsilon^2 . \tag{41}$$

Finally we have

$$\Theta(r,t) = \Theta(r,t_g) + k_\vartheta \bar{\upsilon}(r,t_g)(t-t_g) - \frac{k_\vartheta}{3}\left(\frac{\partial \bar{\upsilon}(r,t_g)}{\partial r}\right)\left(\frac{\partial \upsilon_{gr}}{\partial s}\right)_{t_g} (t-t_g)^3 + \ldots . \tag{42}$$

It is the traveling wave expressed for sufficiently small $\varepsilon$ with 'linear frequency' $\omega_g \equiv k_\vartheta \bar{\upsilon}(r,t_g)$. The rapidity $\left(\partial \upsilon_{gr}/\partial s\right)_{t_g}$ plays a role at the order of $O(\varepsilon^3)$. The comparison of equation (42) with the numerical results is shown in figure 14. The agreement holds for a short interval. However, the perturbative solution breaks down because of the oscillatory structure of zonal flow. Let us take the yellow line ($\rho = 0.9$) in figure 14(a) as the example. The perturbative solution departs from the numerical one around 34.82ms. The expansion to higher order does not help altering such a trend. When we look at figure 13(b) near $\rho = 0.9$, $\partial \bar{\upsilon}(r)/\partial r$ is negative. However, the true argument of zonal flow in equation (39) is not $r$, but $r - r_g(t,t',t_g)$. The negative $r_g$ pushes true argument of zonal flow to $\rho = 0.905$ at 34.82ms. It is precisely the position where $\partial \bar{\upsilon}(r)/\partial r$ reverses sign. It thus alters up-trend to down-trend.

In other words, the fine radial structure of zonal flow in the reaction region breaks up the Taylor expansion based perturbative solution.

It is also very interesting to see in figure 14 that at the late stage of Instanton life, say 35.1ms in (a), the lines marking radial position are in the same order as defined by the blue line for Caviton in figure 11. It implies that the spatial structure of Caviton inherits that generated by the preceding Instanton.

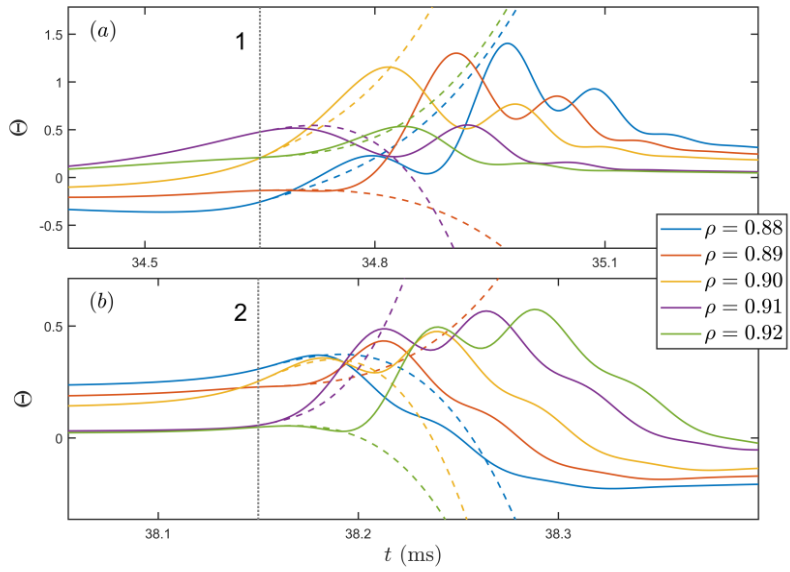

Figure 14. Comparison of analytic results equation (42) to numerical results of figure 10(a) and (b)

Now, we come to the most important information regarding Instanton, the time evolution of Instanton frequency shown in figure 15. It suggests that for GAM frequency in range of 10-20 kHz the turbulence level at $I_m = 10^{-3}$ is adequate to make Instanton resonant to the GAM frequency, since our numerical results have shown that the maximum of Instanton frequency increases with turbulence level (details are neglected). This implies that static Reynolds stress alone does not generate GAM. The causal relationship between GAM excitation and frequency of Instanton will be presented in the next section.

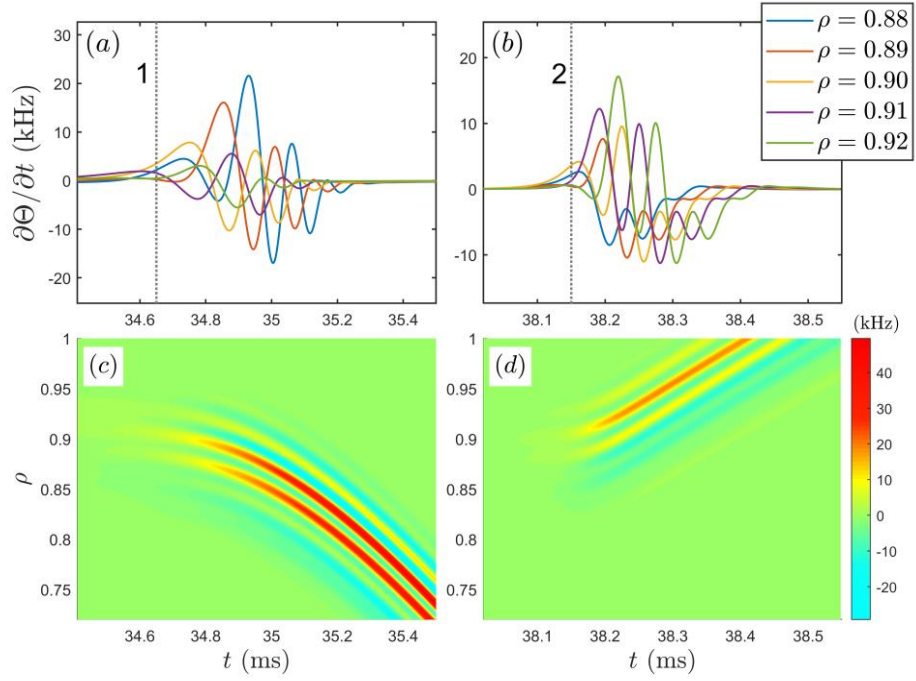

Figure 15. The time evolution of frequency $\partial \Theta / \partial t$ and its spatiotemporal structure in period 1 and 2 in figure 10

## 7. The intermittent excitation of GAM by nonlinear Instanton

In order to demonstrate the intermittent excitation of GAM by nonlinear Instanton it is appropriate to use the zonal flow equation derived by Chakrabarti *et.al.* [29] for EDW. In Appendix C the single equation in terms of zonal flow is given by equation (C9). Apparently, it is the zero-dispersion equation, and zero propagation. It means $D(\tau_e) = 0$ in the dispersion relation in Fourier representation [32]

$$\omega^2 = \omega_G^2 \left[ 1 + D(\tau_e) \rho_i^2 k^2 \right], \quad \omega_G^2 \equiv 2(1+\tau_i)\frac{c_s^2}{R^2}\left(1+\frac{1}{2q^2}\right). \tag{43}$$

It has been shown that in cold ion limit $D(\tau_e)\rho_i^2 \to -\rho_s^2$. With increasing $\tau_i$, $D(\tau_e)$ becomes positive [32,33]. Many derivations based on gyro-kinetics do not have cold ion limit and yield positive $D(\tau_e)$ [32,34]. The generic fluid model can only yield negative $D(\tau_e)$.

However, it may yield positive $D(\tau_e)$ by invoking finite Larmor radius correction [33]. In GAM experiments the GAM wave length is typically much longer than ion Larmor radius $\rho_i$ [35,36], it suggests that the specific value of $D(\tau_e)$ is not so important. Then the GAM zero dispersion is modified to be

$$\frac{\partial^2}{\partial t^2} \to \frac{\partial^2}{\partial t^2}\left(1 + D(\tau_e)\rho_i^2 \frac{\partial^2}{\partial r^2}\right). \tag{44}$$

As a result, equation (C8) becomes

$$\frac{\partial^2}{\partial t^2}\left(1 + D(\tau_e)\rho_i^2 \frac{\partial^2}{\partial r^2}\right)\chi + \frac{(1+\tau_i)}{q^2}\frac{c_s^2}{R^2}\chi - \frac{2}{R}\frac{\partial \bar{\upsilon}}{\partial t} = 0. \tag{45}$$

On the other hand, since the EDW envelope is modulated by $\cos\Theta$, resulting in $\cos^2\Theta$ modulation to Reynolds stress, equation (C7) should be replaced by

$$\frac{\partial \bar{\upsilon}}{\partial t} - \mu \frac{\partial^2 \bar{\upsilon}}{\partial r^2} = -(1+\tau_i)\frac{c_s^2}{R}\chi - \frac{\partial}{\partial r}\left[\Re_{\vartheta,j}(r)\cos^2\Theta\right]. \tag{46}$$

The sound wave equation (45) and the zonal flow equation (46) combined with the envelope equation (38), constitute the basic equation set to illustrate the intermittent excitation of zonal flow by nonlinear Instantons in the next two sub-sections. They are numerically solved with the same parameters in section 6 and $D(\tau_e) = 1$ for illustration.

### 7.1 GAM excitation by nonlinear Instanton for single central rational surface

The time evolution of radial group velocity, zonal flow $\bar{\upsilon}$ and phase $\Theta$ are displayed in figure 16 respectively. Comparing with results in figure 9, one can clearly see that besides the TLFZF defined by equation (C10) in Appendix C, a high frequency packet emerges abruptly around the zero-crossing time of radial group velocity. The close-up temporal structure of Instanton frequency $\partial\Theta/\partial t$ and zonal flow in period 1 and 2 are displayed in figure 17. It is apparent that the high frequency zonal flow arises from the rapid variation of the phase $\Theta$. We have also analyzed results for lower turbulence intensity, e.g., $I_m(r_m) = 10^{-4}$ (details are neglected), there is almost no GAM excited since the maximum of $\partial \Theta / \partial t$ is around 3kHz, far below the local GAM frequency $\omega_G$ =13.64 kHz at $\rho = 0.9$.

The excitation of GAM by Instanton is also illustrated in figure 18 corresponding to period near 2 in figure 17(d). This is the example that outgoing Instanton induces outward propagating GAM. Among others the wave length of GAM is about 1cm. It is at the same order of that observed in experiments [35,36], however, much longer than the ion Larmor radius. It suggests that the spatial structure of Instanton could be more important than the dispersion coefficient in determining the wave length of GAM.

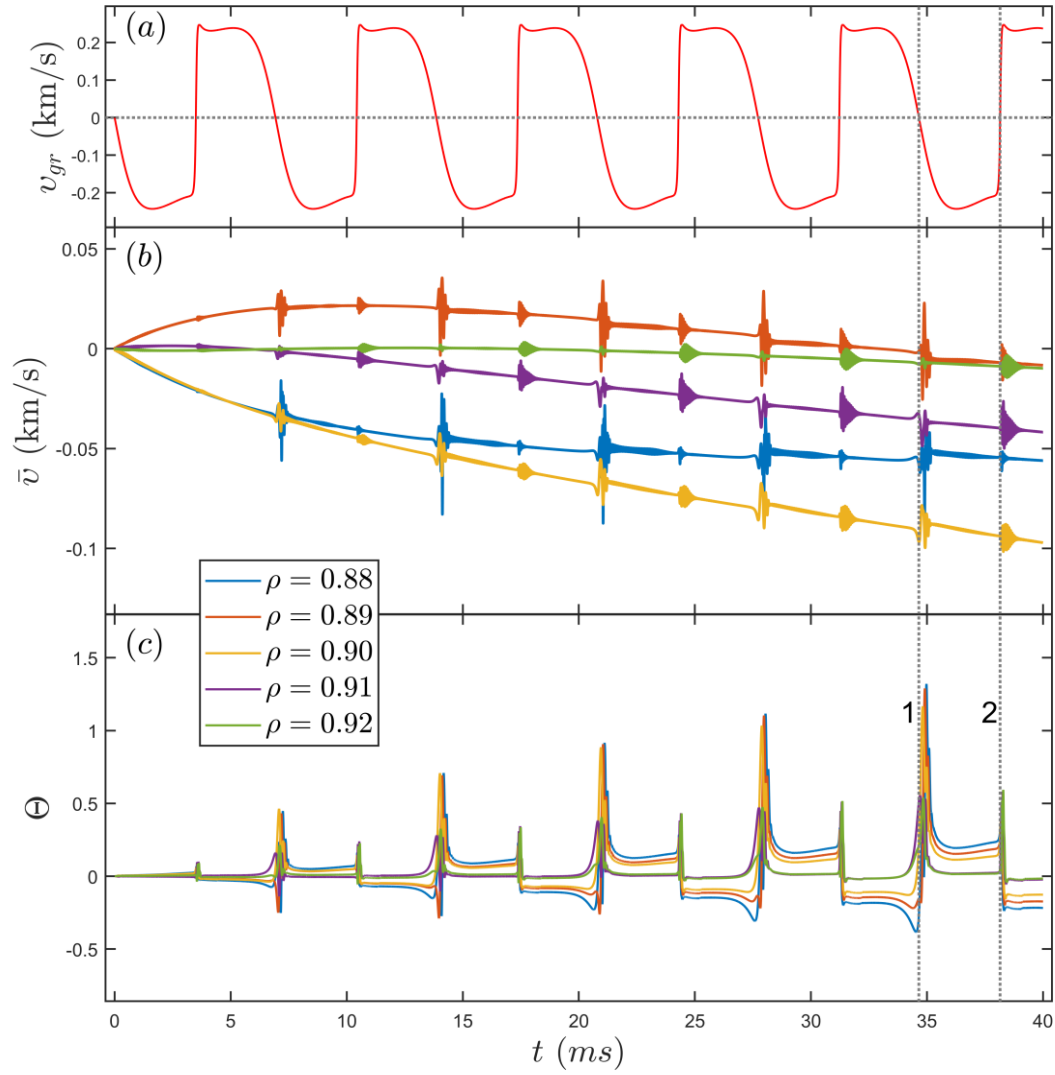

Figure 16. The time evolution of 40ms for 5 distinct spatial positions to display the relationship of zonal flow including GAM, phase spikes and zero-crossing of radial group velocity

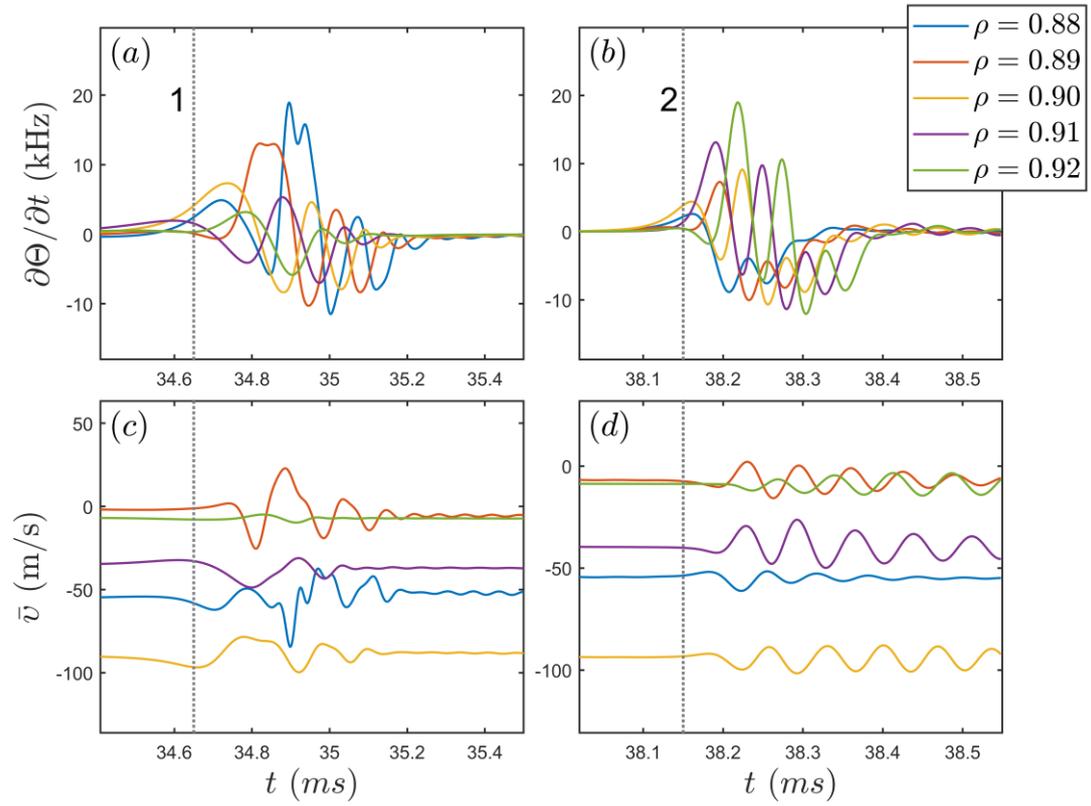

Figure 17. The time evolution of frequency $\partial\Theta/\partial t$ and the close-up of zonal flow in period 1 and 2 in figure 16

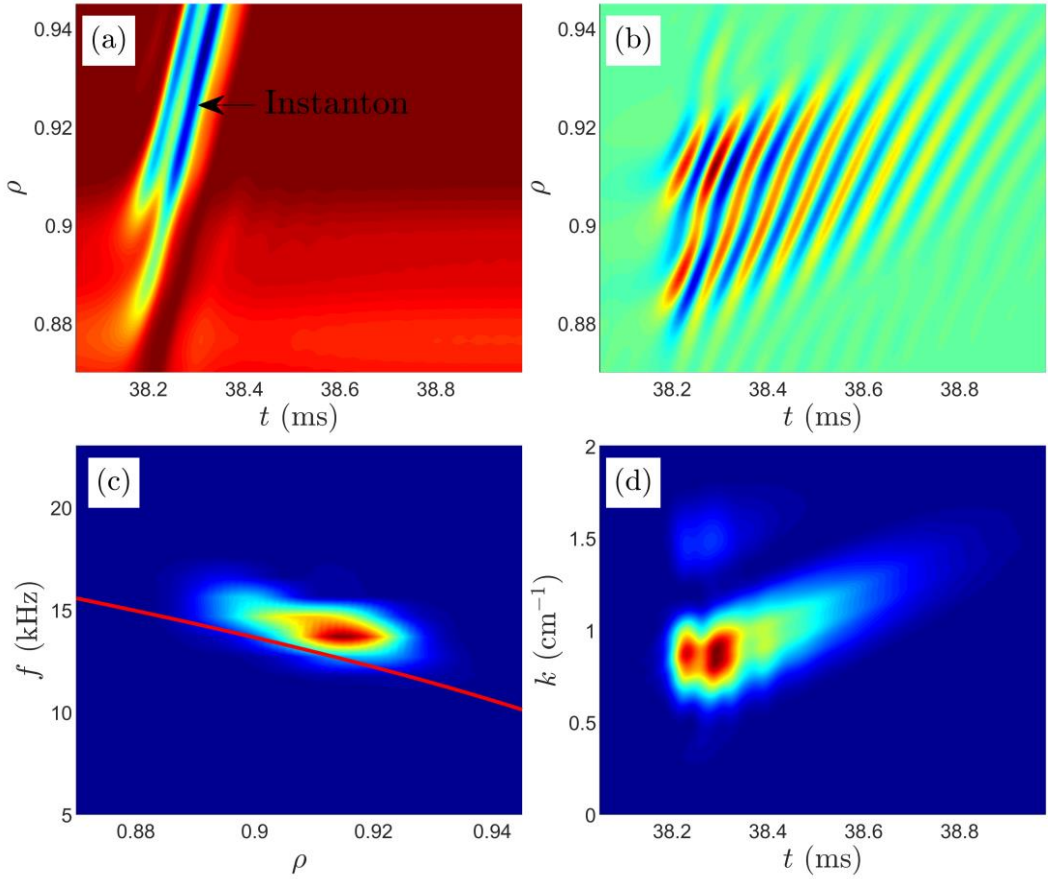

Figure 18. Spatiotemporal structure of (a) Instanton $I_m \propto \cos^2\Theta$ and (b) $\bar{\upsilon}_{\text{GAM}}$ for $D(\tau_e) = 1$ in period 38.05-39ms and $\rho = 0.87 - 0.94$, (c) spectrum versus frequency and spatial position (the red line represents the local GAM frequency $\omega_G$), (d) spectrum versus wave number and time

## 7.2 Nonlinear coupling between multiple central rational surfaces and GAM intermittency

In practice, zonal flow could be driven by nonlinear coupling between multiple central rational surfaces, where the 'coupling' refers to share of the same zonal flow for each individual central rational surface. For EDW in this paper, there are about 5 rational surfaces within the reaction region (where Reynolds stress in figure 6 is not small), all of them contribute to a poloidal

- 38 -

torque $\Re_{\vartheta,j}(r)$ ($j = 0, \pm 1, \pm 2$), resulting in the total torque $\sum_j \Re_{\vartheta,j}(r)\cos^2\Theta_j$. Here $\Theta_j$ represents the contribution from the *j*th central rational surface. The relative phase between different $\Theta_j$ is indefinite, because the initial value of integration along the poloidal characteristic line $d\vartheta/dt = \upsilon_{gy}(\vartheta)/r_j$, could not be pre-determined in $[0, 2\pi]$. For brevity, this is called random phase mixing between multiple central rational surfaces in the reaction region. It results in the indefinite zero-crossing time of radial group velocity for those rational surfaces shown in figure 19(a). As a result, GAMs are triggered asynchronously for different central rational surfaces, forming a rather intricate pattern displayed in figure 19(b). Corresponding frequency-time spectrogram is shown in figure 19(c), very similar to experimental measures of GAM intermittency [25,37-44].

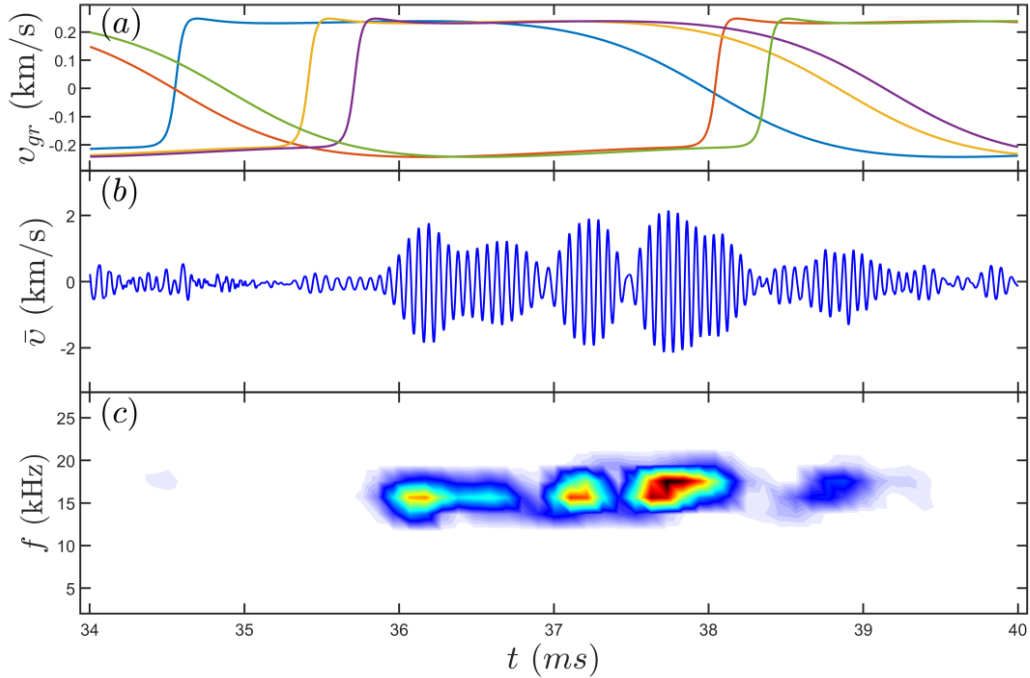

Figure 19. (a) Temporal evolution of radial group velocity pertaining to 5 rational surfaces, (b) zonal flow at $\rho = 0.9$ and (c) corresponding spectrogram

## 8. Summary and on the motion of drift wave envelope

In this paper the full toroidal zonal flow-drift wave system, driven by turbulence associated with an EDW (fluid $\delta_e$-model), is derived and investigated. The model is set for a large aspect ratio tokamak configuration. By making use of derivative expansion, the two-scale system is constructed in section 2. The micro-scale is described by the 2D (toroidal) $\delta_e$-mode, entailed by the eigenmode equation; not only eigenvalue, but also mode structure is solved by making use of 2D ballooning theory, namely WABT. The meso-scale is described by the EDW envelope equation modulated by zonal flow. The EDW mode structure is solved in $(k,\lambda)$ representation first via WABT, then converted into $(x,l)$ representation. This theoretical (asymptotic) solution is further refined by making use of iterative differece method to obtain more accurate numerical solution. The mode structure is used to calculate two significant meso-scale qunatities, Reynolds stress contained in the zonal flow equation as the source term and group velocity contained in the drift wave envelope equation describing characteristics. The caluculation is conducted in spatiotemporal representation for the following reasons. The Reynolds stress contains subttle spatial symmetries in the calculation. The detailed knowledge of time evolution in group velocity is found crucial [8].

The Reynolds stress and group velocity in terms of EDW are then substituded into the meso-scale equations (37) , (38) in section 6 to study salient features of EWD – zonal flow sytem, in particular, the evolution of Caviton-Instanton are analyzed and displayed in graphs. Right after zero-crossing of radial group velocity $\upsilon_{gr}$, the linear traveling wave (linear Instanton) emerges from the standing wave Caviton, as shown by $\Delta\Theta(r,t) = k_g \bar{\upsilon}(r,t_g)(t-t_g)$, then it keeps growing to the nonlinear stage as shown by equation (42). The frequency of Instanton is also growing to its maximum as shown in figure 15(a) and (b). Depending on the turbulence level $I_m$, higher $I_m$ would yield higher maximum of frequency. The intermittent excitation of GAM by nonlinear Instanton has been unambiguously demonstrated in section 7. It would, now, be appropriate to summarize the new physics extracted and highlighted in this paper:

(1) The EDW envelope is not always in the standing wave (Caviton) phase. Interactions with the zonal flow may lead to the sudden emergence of the short-lived traveling wave in the radial direction (Instanton). This Caviton to Instanton transition, occurring upon the zero-crossing of radial group velocity, is a major event in the evolution of the system.

(2) The Caviton could be positive (bump) or negative (trough). The Caviton to Instanton transition results in sign reversal, the sign of next Caviton becomes opposite to that of the previous one. The spatial structure of Caviton inherits the radial integration over zonal flow; its sign is determined by the sign of radial group velocity.

(3) Similarly, there are two types of Instanton, ingoing and outgoing. The ingoing (outgoing) is induced by downward (upward) zero-crossing. It is also associated with the sign of Caviton. The positive (negative) Caviton comes after ingoing (outgoing) Instanton.

(4) The initial Instanton is a linear traveling wave in radial direction, and, then, it evolves to a nonlinear traveling wave. The amplitude and frequency in the nonlinear stage increases with time initially, it reaches a peak, then oscillates due to the fine radial structure of zonal flow; finally, it falls off because the Instanton travels out of the system.

(5) The maximal frequency of Instanton depends on the turbulence level. For the parameters in this paper $I_m = 10^{-3}$ ($I_m = 10^{-4}$), the maximal frequency of Instanton is 20 kHz (3 kHz). Since 20 kHz is high enough to the GAM frequency $\omega_G$ (13.64 kHz), the GAM is excited by resonance.

(6) The lifetime of resonance - excited GAM is not long enough to form the 'WKB eigenmode' [45]. The eigenmode formation would require several times of back and forth reflection between the two boundaries.

(7) The resonance - excited GAM propagates radially with a wave length much longer than the Larmor radius. This is consistent with observations in experiment [35,36].

(8) GAM intermittency follows from the nonlinear coupling between multiple central rational surfaces. The integration along the poloidal characteristics brings in the indefinite initial values pertaining to each central rational surface. It can be called 'random phase mixing' between several to decades of central rational surfaces, because the relative phases between the components participating in the coupling are indefinite. For each discharge they are different. Therefore, the pattern as shown in figure 19(c) is not repeatable.

Finally, we would like to point out that while the theory of GAM associated with micro-turbulence in ion direction (ITG) already exists in literature [46], including intermittent excitation and propagation as well, it seems still premature at this stage to analyze to detail for comparison. For the time being we just say that the qualitative features of GAM induced by either ITG or EDW are essentially similar. In future we prefer to fulfill the task from category to category. For example in discussing the wave length of GAM, both results of ITG and EDW will be presented with comparisons.

## Appendix A Derivation of $\delta_e$-model in $(x,l)$ representation

The $\delta_e$-model equation (the zeroth order equation (9) derived in section 2) is

$$\left[\left(1+\frac{\hat{\omega}_{*i}}{\omega}\right)\rho_s^2\nabla_\perp^2 - (1-i\delta_e) + \frac{\hat{\omega}_{*e}}{\omega} - 2(1+\delta_\tau)\frac{\hat{\omega}_{de}}{\omega} - (1+\delta_\tau)\frac{c_s^2}{\omega^2}\nabla_\parallel^2\right]\varphi(\tilde{r}) = 0. \quad (A1)$$

In the $(x,l)$ representation, substituting equation (13) into equation (A1), the differential operators are

$$\nabla_\perp^2 \varphi \to k_g^2\left[\hat{s}^2\frac{\partial^2}{\partial x^2} - \left(1+\frac{l}{m}\right)^2\right]\varphi_l(x),$$

$$\nabla_\parallel \varphi = \frac{1}{qR}\left[\frac{\partial}{\partial\vartheta} + q(r)\frac{\partial}{\partial\zeta}\right]\varphi \to \frac{i}{qR}(x-l)\varphi_l(x). \quad (A2)$$

The three frequency operators are

$$\hat{\omega}_{*e,i}\varphi \to \omega_{*e,i}\left(1+\frac{l}{m}\right)\varphi_l(x), \quad \hat{\omega}_{de}\varphi \to \omega_{de}\left[i\hat{s}\sin\vartheta\frac{\partial}{\partial x} + \cos\vartheta\left(1+\frac{l}{m}\right)\right]\varphi_l(x), \quad (A3)$$

Use is made of the linear density profile, $f_n(x) \equiv n^{(0)}/n_j = 1-(r_j/L_n)(x/m\hat{s})$, temperature profile $f_{T_s}(x) \equiv T_s^{(0)}/T_{s,j} = 1-(r_j/L_{T_s})(x/m\hat{s})$ ( $s=i,e$ ), ion pressure profile $f_{P_i}(x) \equiv P_i^{(0)}/P_{i,j} = 1-(r_j/L_{P_i})(x/m\hat{s})$, $P_i^{(0)} \equiv n^{(0)}T_i^{(0)}$. The density, electron temperature and ion pressure gradient length: $L_n \equiv \left(d\ln n^{(0)}/dr\right)_{r_j}^{-1}$, $L_{T_s} \equiv \left(d\ln T_s^{(0)}/dr\right)_{r_j}^{-1}$, $L_{P_i} \equiv \left(d\ln P_i^{(0)}/dr\right)_{r_j}^{-1}$. $\rho_s^2 = \rho_{s,j}^2 f_{T_e}(x)$ and $c_s^2 = c_{s,j}^2 f_{T_e}(x)$ also depend on electron temperature, subscript $j$ denotes equilibrium quantities on rational surface $r_j$. The sideband $l/m$ and slow variation of equilibrium profile in radial direction violate the translational invariance.

After introducing the dimensionless parameters defined below equation (14), the equation for $\varphi_l(x)$ is

$$\hat{k}_g^2 \frac{\hat{\omega}}{\hat{\omega}_s} \left[1 + \frac{\bar{\eta}_i}{\hat{\omega}} \frac{f_{T_i}(x)}{f_{P_i}(x)}\left(1+\frac{l}{m}\right)\right]\left[\hat{s}^2 \frac{\partial^2}{\partial x^2} - \left(1+\frac{l}{m}\right)^2\right]\varphi_l$$

$$-\frac{1}{\hat{\omega}_s}\left[(1-i\delta_e)\frac{\hat{\omega}}{f_{T_e}(x)} - \frac{1}{f_n(x)}\left(1+\frac{l}{m}\right)\right]\varphi_l + (1+\delta_\tau)\frac{\hat{\omega}_s}{\hat{\omega}}(x-l)^2 \varphi_l \quad \text{(A4)}$$

$$-(1+\delta_\tau)\frac{\hat{\omega}_{de}}{\hat{\omega}_s}\left[\hat{s}\frac{\partial}{\partial x}(\varphi_{l+1}-\varphi_{l-1}) + \left(1+\frac{l}{m}\right)(\varphi_{l+1}+\varphi_{l-1})\right] = 0$$

To simplify the derivation, we assume derivatives are applied only on the wave functions, not on the equilibrium quantities, and the TSB terms $x$ in the equilibrium profile can be replaced by $l$, $f(x) \approx f(l)$. Since translational invariance is the leading symmetry, all TSB terms are small enough to be retained up to the second order $O(l^2/m^2)$. Then the $\delta_e$-model expanded to include TSB terms in the $(x,l)$ representation is obtained as equation (14).

## Appendix B The conditions for valid WABT

Let us starting with equation (23) and introducing the transform

$$\psi(\lambda) \equiv \Phi(\lambda)\exp\left[-\frac{1}{2}\int^\lambda d\lambda' P(\lambda')\right]. \quad \text{(B1)}$$

Then the equation for $\Phi(\lambda)$ follows from equation (23) and equation (B1)

$$\frac{d^2\Phi}{d\lambda^2} + \left[Q(\lambda) - \frac{P^2}{4} - \frac{1}{2}\frac{dP}{d\lambda}\right]\Phi = 0. \quad \text{(B2)}$$

In the large $n$ limit

$$P(\lambda) \approx in\frac{\bar{L}_1^{(0)}(\lambda;\hat{\omega})}{\bar{L}_2^{(0)}(\lambda;\hat{\omega})} \to in\frac{\bar{L}_1}{\bar{L}_2},$$

$$Q(\lambda) \equiv \frac{n^2}{\bar{L}_2^{(0)}(\lambda;\hat{\omega})}\left[\Omega(\lambda) - \Omega(\hat{\omega})\right] \to \frac{n^2}{\bar{L}_2}\left[\Omega(\lambda) - \Omega(\hat{\omega})\right]. \quad \text{(B3)}$$

In the second step we neglect the $\lambda$ dependence in $\bar{L}_1^{(0)}$ and $\bar{L}_2^{(0)}$, i.e. $\bar{L}_1$ and $\bar{L}_2$ are constants. Further approximation is

$$\Omega(\lambda) \approx \Omega_0 + \Omega_1 \cos\lambda \approx \Omega_0 + \Omega_1\left(1 - \frac{\lambda^2}{2}\right). \tag{B4}$$

Then we obtain the following analytic solution for eigenvalue and wave function

$$\Omega(\hat{\omega}) = \Omega_0 + \Omega_1 + \frac{\overline{L}_1^2}{4\overline{\overline{L}}_2}, \tag{B5}$$

$$\psi(\lambda) = \Phi(\lambda)\exp\left(-in\frac{\overline{L}_1}{2\overline{\overline{L}}_2}\lambda\right), \quad \Phi(\lambda) = \exp\left(-\sqrt{\frac{\Omega_1}{2\overline{\overline{L}}_2}}\frac{n\lambda^2}{2}\right). \tag{B6}$$

This is the WABT solution. For $\overline{L}_1 = 0$ it reduces to horizontal ballooning theory (HBT) [15]. It exhibits fast variation of FPD with Gaussian width $\Delta\lambda \sim 1/\sqrt{n}$. Therefore, the true small parameter in large $n$ limit is $\varepsilon_B \equiv (1/n)(d/d\lambda) \sim 1/\sqrt{n}$. For finite $\overline{L}_1 \neq 0$ (WABT) the exponential factor in the first equation of equation (B6) introduces a fast variation unless $\Xi \equiv \overline{L}_1 / 2\overline{\overline{L}}_2 \ll 1$, otherwise the higher order derivative would contribute, and WABT becomes invalid. In case such a parameter regime comes true, the observed flavor could be vertical ballooning theory (VBT) [15].

Practically, two functions in equation (B6) yields two criterions, because the general criterion $\varepsilon_B \ll 1$ consists of two conditions. For very small $\Xi$, $(1/n)(d/d\lambda) \approx \sqrt{\Omega_1/2n\overline{\overline{L}}_2}$, the same as HBT. However, there exists restriction on $\Xi$. The estimate for $\Delta\lambda \approx 1/\sqrt{n}$ leads to $\Xi\sqrt{n} \ll \pi/2$. Both should be satisfied for validity of WABT.

## Appendix C The zonal flow equation for electron drift wave

The zonal flow equation for EDW containing both the low-high frequency branches has been derived in [29]. Since that version is written in Fourier representation, the equations have to be cast into the spatiotemporal representation before being applied to this paper. We shall use the same basic equations, equations (21-23) of [29]. They are the governing equations for GAM in dimensional form provided validity of scale separation. The three equations are the continuity equation

$$\frac{\partial \bar{n}_G}{\partial t} + \nabla_\parallel \bar{u}_G - \frac{2\rho_s c_s}{R} \sin\vartheta \frac{\partial \bar{\varphi}}{\partial r} = 0 \,, \tag{C1}$$

the vorticity equation

$$\frac{\partial}{\partial t} \rho_s^2 \nabla_\perp^2 \bar{\varphi} + (1+\tau_i)\frac{2\rho_s c_s}{R}\sin\vartheta \frac{\partial \bar{n}_G}{\partial r} - \mu\rho_s^2 \frac{\partial^4 \bar{\varphi}}{\partial r^4} = \rho_s^3 c_s \frac{\partial^2}{\partial r^2}\left\langle \frac{\partial \varphi}{r\partial \vartheta}\frac{\partial \varphi}{\partial r}\right\rangle_{en}, \tag{C2}$$

and the parallel momentum equation

$$\frac{\partial \bar{u}_G}{\partial t} = -c_s^2(1+\tau_i)\nabla_\parallel \bar{n}_G \,, \tag{C3}$$

where $\bar{n}_G$ and $\bar{u}_G$ are density and parallel velocity fluctuations in meso-scale associated with GAM respectively, both are axisymmetric with $m=1$, $n=0$, $\bar{\varphi}$ is the zonal electrostatic potential, with $m=0$, $n=0$. The term on right hand side of (C2) is the ensemble average of Reynolds stress in spatiotemporal representation [28], where $\varphi$ is the EDW eigenmode defined in equation (13). This term represents the process that two microscale electrostatic potentials (like drift waves) $\varphi$ are annihilated into (poloidal) torque in meso-scale, known as 'three wave interaction' in literature. What will be different from [29] is the vorticity equation, we retain the perpendicular viscosity term which is important for numerical computation.

Substituting equation (C3) into equation (C1) eliminates $\bar{u}_G$,

$$\frac{\partial^2 \bar{n}_G}{\partial t^2} - (1+\tau_i)c_s^2 \nabla_\parallel^2 \bar{n}_G - \frac{2\rho_s c_s}{R}\sin\vartheta \frac{\partial}{\partial t}\frac{\partial \bar{\varphi}}{\partial r} = 0 \,. \tag{C4}$$

This equation mathematically suggests that the only relevant component to the density response in meso-scale is the sinusoidal one. Therefore, it leads us to the solution $\bar{n}_G = \chi(r,t)\sin\vartheta$. In the toroidal coordinates $\nabla_\parallel \to (1/qR)\partial/\partial\vartheta$, equation (C4) becomes

$$\frac{\partial^2 \chi}{\partial t^2} + \frac{(1+\tau_i)}{q^2}\frac{c_s^2}{R^2}\chi - \frac{2\rho_s c_s}{R}\frac{\partial}{\partial t}\frac{\partial \bar{\varphi}}{\partial r} = 0 \,. \tag{C5}$$

This is the sinusoidal sound wave coupled with zonal flow.

The zonal flow equation is obtained by making use of poloidal averaging (in meso-scale) over the vorticity equation. Then we have

$$\rho_s^2 \frac{\partial^2}{\partial r^2}\frac{\partial \overline{\varphi}}{\partial t} + (1+\tau_i)\frac{\rho_s c_s}{R}\frac{\partial \chi}{\partial r} - \mu \rho_s^2 \frac{\partial^4 \overline{\varphi}}{\partial r^4} = \rho_s^3 c_s \frac{\partial^2}{\partial r^2}\left(\oint d\vartheta \frac{\partial \varphi}{r\partial \vartheta}\frac{\partial \varphi}{\partial r}\right). \tag{C6}$$

The quantity behind the second order radial derivative on the r.h.s. of equation (C6) is the well-known Reynolds stress [28]. By defining the zonal flow speed $\overline{\upsilon} \equiv \rho_s c_s \partial \overline{\varphi}/\partial r$, equation (C2) becomes

$$\frac{\partial \overline{\upsilon}}{\partial t} - \mu \frac{\partial^2 \overline{\upsilon}}{\partial r^2} = -(1+\tau_i)\frac{c_s^2}{R}\chi - \frac{\partial}{\partial r}\Re_{\vartheta,j}(r), \tag{C7}$$

where $\Re_{\vartheta,j}(r)$ is defined by equation (33), and equation (C5) becomes

$$\frac{\partial^2 \chi}{\partial t^2} + \frac{(1+\tau_i)}{q^2}\frac{c_s^2}{R^2}\chi - \frac{2}{R}\frac{\partial \overline{\upsilon}}{\partial t} = 0. \tag{C8}$$

Through eliminating $\chi$ in equations (C7), (C8) we get the single equation for zonal flow as

$$\left[\frac{\partial^2}{\partial t^2} + 2(1+\tau_i)\left(1+\frac{1}{2q^2}\right)\frac{c_s^2}{R^2}\right]\left[\frac{\partial \overline{\upsilon}}{\partial t} - \mu \frac{\partial^2 \overline{\upsilon}}{\partial r^2} + \frac{\partial}{\partial r}\Re_{\vartheta,j}(r)\right] =$$

$$2(1+\tau_i)\frac{c_s^2}{R^2}\left[-\mu \frac{\partial^2 \overline{\upsilon}}{\partial r^2} + \frac{\partial}{\partial r}\Re_{\vartheta,j}(r)\right]. \tag{C9}$$

In low frequency limit, the temporal derivative in the first term can be neglected, which yields TLFZF equation with the inertia $1+2q^2$, instead of 1 for SLFZF,

$$(1+2q^2)\frac{\partial \overline{\upsilon}}{\partial t} - \mu \frac{\partial^2 \overline{\upsilon}}{\partial r^2} + \frac{\partial}{\partial r}\Re_{\vartheta,j}(r) = 0. \tag{C10}$$

The dispersion of high frequency branch can be obtained from the left hand side of equation (C9)

$$\omega^2 = \omega_G^2 \equiv 2(1+\tau_i)\frac{c_s^2}{R^2}\left(1+\frac{1}{2q^2}\right), \tag{C11}$$

the same as Winsor dispersion relation [47].

**Appendix D Numerical methods for the dimensionless zonal flow equation set**

In this Appendix the normalization and numerical methods for the zonal flow equation set are introduced in detail. The zonal flow speed is normalized to $\bar{\upsilon}_z \equiv \rho_s c_s k_\vartheta \hat{s}\sqrt{I_m(r_m)}$ ($\bar{\upsilon} = V\bar{\upsilon}_z$), time is normalized to $\omega_Z \equiv \bar{\upsilon}_z k_\vartheta$ ($t\omega_Z = \tau$), the dimensionless micro-radius is defined as $x = k_\vartheta \hat{s}(r-r_j)$. The zonal flow Reynolds number [8] is defined as $R_z \equiv \omega_Z / \mu k_\vartheta^2 \hat{s}^2$ ($\mu = 3a_\mu \nu_{ii}\rho_i^2/10$), and the dimensionless Reynolds stress defined in section 5 is $R(x) \equiv \Re_{\vartheta,j}(r)/\rho_s^2 c_s^2 k_\vartheta^2 \hat{s} I_m(r_m)$. Then we define the dimensionless sinusoidal component of sound wave to be $\mathrm{X} \equiv k_\vartheta R\chi$. By introducing the sound frequency $\varpi_G^2 \equiv (1+\tau_i)c_s^2/q^2R^2$ varying in equilibrium scale, and $\hat{\varpi}_G \equiv \varpi_G/\omega_Z$, the sound wave equation (45) and the zonal flow equation (46) combining with the envelope equation (38) can be cast into the dimensionless form

$$\frac{\partial^2}{\partial \tau^2}\left(1+\delta^2 \frac{\partial^2}{\partial x^2}\right)\mathrm{X} + f_{T_e}(x)\hat{\varpi}_G^2 \mathrm{X} - 2\frac{\partial V}{\partial \tau} = 0, \tag{D1}$$

$$\frac{\partial V}{\partial \tau} - \frac{1}{R_Z}\frac{\partial^2 V}{\partial x^2} = -f_{T_e}(x)q^2\hat{\varpi}_G^2 \mathrm{X} - \frac{\partial}{\partial x}\left[R(x)\cos^2\Theta\right], \tag{D2}$$

$$\Theta(x,\tau) = \int_0^\tau d\tau' V\left(x - \hat{s}\int_{\tau'}^\tau d\sigma \hat{\upsilon}_{gr}(\sigma), \tau'\right), \tag{D3}$$

where $\hat{\upsilon}_{gr} \equiv \upsilon_{gr}/\bar{\upsilon}_z$, $\delta \equiv \rho_i k_\vartheta \hat{s}\sqrt{D(\tau_e)}$, $f_{T_e}(x)$ describes the (linear) electron temperature profile which has been defined in Appendix A, and is important for GAM propagation [45].

The solution of the initial value problem is worked out for the initial conditions: $V(x,0)=0$, $\Theta(x,0)=0$ and $\mathrm{X}(x,0)=0$. Since the phase modulation of drift wave envelope is significant only inside the reaction region, the boundary condition for drift wave envelope is not important; the latter is relevant only in the Instanton phase that has a vanishing coupling to zonal flow outside reaction region. However, the signal of phase function $\Theta$ could propagate to a place far away from the reaction region denoted by $x_{\pm\infty}$, where the cut-off has been introduced ($\Theta(x_{\pm\infty},\tau)=0$) [8]. In computation $x_{-\infty}$ and $x_{+\infty}$ are set to be $\rho=0.75$ and $\rho=1$ respectively.

The boundary condition for the zonal flow equation set has to be set up differently with respect to left and right side in contrast to [8], since the data of GAM are measured not too far away from plasma edge, and edge effects on GAM could be important. On the right side, the zero Dirichlet (reflecting) boundary condition is chosen at the plasma edge for $V(x_{\text{edge}},\tau)=0$, $\chi(x_{\text{edge}},\tau)=0$, where $x_{\text{edge}}$ denotes the position of plasma edge ($\rho=1$), for the reason that GAM cannot propagate outside the last closed magnetic surface. On the left side, an absorptive boundary condition is set up since GAM suffers from Landau damping in low $q$ region [48].

The dimensionless set of equations (D1-D3), combined with the assumed initial and boundary conditions, constitutes a well-posed initial value problem, which is solved by making use of the finite difference methods, where the spatiotemporal grids are discretized as $(r_k,t_m)$, $r_k \equiv r_1 + k\cdot\Delta r$, $t_m \equiv m\cdot\Delta t$, k=0,1,...,K and $m=0,1,...,M$ are integers. In this paper we choose $K=600$, $M=20000$, $\Delta r=(r_2-r_1)/K$, $\Delta t=2\mu s$, $r_1$ and $r_2$ are the position of the left and right boundary, corresponding to $\rho=0.75$ and $\rho=1$ respectively. The dimensionless spatiotemporal step sizes are $\Delta x \equiv |k_\vartheta \hat{s}|\cdot\Delta r$ and $\Delta\tau \equiv \omega_Z \cdot \Delta t$. equations (D1-D2) are solved by making use of the 2$^{\text{nd}}$ order Crank-Nicolson method [49]. The envelope equation (D3) is directly integrated as shown in Appendix C of [8].

As for the nonlinear coupling between multiple central rational surfaces, equations (D2-D3) should be replaced by

$$\frac{\partial V}{\partial \tau} - \frac{1}{R_Z}\frac{\partial^2 V}{\partial x^2} = -f_{T_e}(x)q^2\hat{\varpi}_G^2 X - \frac{\partial}{\partial x}\left[\sum_{j=-j_{max}}^{j_{max}} R(x+j)\cos^2\Theta_j\right], \tag{D4}$$

$$\Theta_j(x,\tau) = \int_0^\tau d\tau' V\left(x+j-\hat{s}\int_{\tau'}^\tau d\sigma \hat{\upsilon}_{gr}(\tau_j+\sigma),\tau'\right), \tag{D5}$$

where $\tau_j$ represents the arbitrary initial phase of each rational surface. It is reasonable to assume the mapping between the time and poloidal angle $\vartheta(\tau_j)$ to be a random number in $[0,2\pi]$. Since the interval of adjacent rational surface is $\Delta x = 1$, the radial argument $x$ is replaced by $x+j$. $j_{max} = 2$ is set for illustration in section 7.

## Acknowledgments

The present work was supported in part by the National MCF Energy R&D Program under Grant Nos. 2018YFE0311200 and 2017YFE0301204, the Natural Science Foundation of China under Grant Nos. No. U1967206 and 11975231, National Natural Science Foundation of China (NSFC-11805203, 11775222), Key Research Program of Frontier Science CAS (QYZDB-SSW-SYS004) and the U.S. Dept. of Energy Grant No. DE -FG02-04ER-54742.

## References

[1] Diamond P H, Itoh S-I, Itoh K, Hahm T S 2005 Plasma Phys. Control. Fusion **47** R35-R161

[2] Fujisawa A 2009 Nucl. Fusion **49** 013001

[3] Miki K, Kishimoto Y, Miyato N and Li J Q 2007 Phys. Rev. Lett. **99** 145003

[4] Miki K, Kishimoto Y, Li J Q and Miyato N 2008 Phys. Plasmas **15** 052309

[5] Storelli A, Vermare L, Hennequin P, Gurcan O D, Dif-Pradalier G, Sarazin Y, Garbet X, Gorler T, Singh R, Morel P *et al* 2015 Phys. Plasmas **22** 062508

[6] Gurchenko A D, Gusakov E Z, Niskala P, Altukhov A B, Esipov L A, Kiviniemi T P, Korpilo T,

Kouprienko D V, Lashkul S I, Leerink S, Perevalov A A and Irzak M A 2016 Plasma Phys. Control. Fusion **58** 044002

[7] Askinazi L G, Belokurov A A, Bulanin V V, Gurchenko A D, Gusakov E Z, Kiviniemi T P, Lebedev S V, Kornev V A, Korpilo T, Krikunov S V, Leerink S, Machielsen M, Niskala P, Petrov A V, Tukachinsky A S, Yashin A Yu and Zhubr N A 2017 Plasma Phys. Control. Fusion **59** 014037

[8] Zhang Y Z, Liu Z Y, Xie T, Mahajan S M and Liu J 2017 Physics of Plasmas **24** 122304

[9] Tang W M 1978 Nucl. Fusion **18** 1089

[10] Tsang K T 1977 Nucl. Fusion **17** 261

[11] Horton W, Estes Jr R D, Kawk H and Choi D I 1978 Phys. Fluids **21** 1366

[12] Chen L and Cheng C Z 1980 Phys. Fluids **23** 2242

[13] Zhang Y Z, Mahajan S M and Zhang X D 1992 Phys. Fluids B **4** 2729

[14] Zhang Y Z and Mahajan S M 1992 Phys. Fluids B **4** 207

[15] Xie T, Qin H, Zhang Y Z and Mahajan S M 2016 Phys. Plasmas **23** 042514

[16] Pearlstein L D and Berk H L 1969 Phys. Rev. Lett. **23** 220

[17] Ross D W and Mahajan S M 1978 Phys. Rev. Lett **40** 324

[18] Tsang K T, Catto P J, Whitson J C and Smith J 1978 Phys. Rev. Lett. **40** 327

[19] Cheng C Z and Chen L 1980 Phys. Fluids **23** 9

[20] Johnson R S 2005 Singular Perturbation Theory (Springer, New York) Chapter 4

[21] Zhang Y Z and Mahajan S M 1991 Phys. Lett. A **157** 133

[22] Dewar R L, Zhang Y Z and Mahajan S M 1995 Phys. Rev. Lett. **74** 4563

[23] Zhang Y Z, Xie T 2013 Nucl. Fusion & Plasma Phys. **33** 193 (in Chinese with English abstract)

[24] Lee Y C and Van Dam J W September 1977 in proceedings of the Finite Beta Theory Workshop, Varenna Summer School of Plasma Physics, Varenna, Italy, edited by Coppi B and Sadowski B (U.S. Dept. of Energy, Office of Fusion Energy 1979 Washington DC) CONF-7709167 p.93

[25] Hillesheim J C, Peebles W A, Carter T A, Schmitz L and Rhodes T L 2012 Phys. Plasmas **19**, 022301

[26] Sewell G 2005 Computational Methods of Linear Algebra, 2rd edition (A John Wiley & Sons, Inc., Hoboken, NJ) Sec.3.5 p.12

[27] Xie T, Zhang Y Z, Mahajan S M, Liu Z Y and Hongda He 2016 Phys. Plasmas **23** 102313

[28] Zhang Y Z and Mahajan S M 1995 Phys. Plasma **2** 11

[29] Chakrabarti N, Singh R, Kaw P K, Guzdar P N 2007 Phys. Plasmas **14** 052308

[30] Hirshman S P and Sigmar D J 1981 Nucl. Fusion **21** 1079

[31] Shaing K C, Ida K and Sabbagh S A 2015 Nucl. Fusion **55** 125001

[32] Smolyakov A I, Bashir M F, Elfimov A G, Yagi M, and Miyato N 2016 Plasma Physics Reports **42** 407

[33] Hager R and Hallatschek K 2009 Phys. Plasmas **16** 072503

[34] Palermo F, Poli E, Bottino A, Biancalani A, Conway G D, and Scott B 2017 Phys. Plasmas **24**, 072503

[35] Liu A D, Lan T, Yu C X, Zhang W, Zhao H L, Kong D F, Chang J F and Wan B N 2010 Plasma Phys. Control. Fusion **52** 085004

[36] Hillesheim J C, Peebles W A, Rhodes T L, Schmitz L, White A E and Cater T A 2010 Review of Scientific Instruments **81** 10D907

[37] Conway G D, Scott B, Schirmer J, Reich M, Kendl A and the ASDEX Upgrade Team 2005 Plasma Phys. Controlled Fusion **47** 1165

[38] Melnikov A V, Vershkov V A, Eliseev L G, Grashin S A, Gudozhnik A V, Krupnik L I, Lysenko S E, Mavrin V A, Perfilov S V, Shelukhin D A *et al* 2006 Plasma Phys. Controlled Fusion **48** S87

[39] Ido T, Miura Y, Hoshino K, Kamiya K, Hamada Y, Nishizawa A, Kawasumi Y, Ogawa H, Nagashima Y, Shinohara K, Kusama Y, JFT-2M group 2006 Nucl. Fusion **46** 512

[40] Cheng J, Yan L W, Zhao K J, Dong J Q, Hong W Y, Qian J, Yang Q W, Ding X T, Duan X R and Liu Y 2009 Nucl. Fusion **49**, 085030

[41] Geng K N, Kong D F, Liu A D, Lan T, Yu C X, Zhao H L, Yan L W, Cheng J, Zhao K J, Dong J Q *et al* 2018 Phys. Plasmas **25** 012317

[42] Silva C, Arnoux G, Groth M, Hidalgo C, Marsen S and JET-EFDA Contributors 2013 Plasma Phys. Controlled Fusion **55** 025001

[43] Silva C, Hillesheim J C, Hidalgo C, Belonohy E, Delabie E, Gil L, Maggi C F, Meneses L, Solano E, Tsalas M and JET Contributors 2016 Nucl. Fusion **56** 106026

[44] Zhou C, Liu A D, Liu Z Y, Wang M Y, Xi F, Zhang J, Ji J X, Fan H R, Shi T H, Liu H Q *et al* 2018 Nucl. Fusion **58**, 106009

[45] Itoh K, Itoh S-I, Diamond P H, Fujisawa A, Yagi M, Watari T, Nagashima Y and Fukuyama A 2006 *Plasma Fusion Res.* **1** 037

[46] Liu Z Y, Zhang Y Z, Mahajan S M, Liu Adi, Xie T, Zhou C, Lan T, Xie J L, Li H, Zhuang G and Liu W D 2021 Plasma Sci. Technol. **23** 035101

[47] Winsor N, Johnson J L and Dawson J M 1968 Phys. of Fluids **11** 2448

[48] Novakovskii S V, Liu C S, Sagdeev R Z and Rosenbluth M N 1997 Phys. Plasmas **4** 4272

[49] Press W H, Teukolsky S A, Vetterling W T and Flannery B P 2007 *Numerical Recipes - The Art of Scientific Computing* 3rd ed. (Cambridge New York)